\documentclass[letterpaper,11pt]{article}
\pdfoutput=1 % if your are submitting a pdflatex (i.e. if you have
\usepackage{jheppub} % for details on the use of the package, please
\usepackage{soul}
\usepackage{tikz}
\usepackage{pgfplots}
\usepackage{subcaption}
\usetikzlibrary{calc,decorations.pathmorphing,shapes}
\usepackage{standalone}
\usepackage[T1]{fontenc} % if needed

\title{\boldmath Link-area commutators in AdS${}_3$ area-networks }

\author[a]{Jesse Held,} 
\author[a]{Molly Kaplan,}
\author[a]{Donald Marolf,}
\author[b,c]{Jie-qiang Wu}
\affiliation[a]{Department of Physics, University of California, Santa Barbara, CA 93106, USA}
\affiliation[b]{CAS Key Laboratory of Theoretical Physics, Institute of Theoretical Physics, Chinese Academy of Sciences, Beijing 100190, China}
\affiliation[c]{School of Physical Sciences, University of Chinese Academy of Sciences, Beijing 100049, China}

\emailAdd{jheld@ucsb.edu}
\emailAdd{mekaplan@ucsb.edu}
\emailAdd{marolf@ucsb.edu}
\emailAdd{jieqiangwu@itp.ac.cn}

\abstract{Random tensor networks (RTNs) have proved to be fruitful tools for modelling the AdS/CFT correspondence. Due to their flat entanglement spectra, when discussing a given boundary region $R$ and its complement $\bar R$, standard RTNs are most analogous to fixed-area states of the bulk quantum gravity theory,  in which quantum fluctuations have been suppressed for the area of the corresponding HRT surface.   However, such RTNs have flat entanglement spectra for all choices of $R, \bar R,$ while quantum fluctuations of multiple HRT-areas can be suppressed only when the corresponding HRT-area operators mutually commute. We probe the severity of such obstructions in pure AdS$_3$ Einstein-Hilbert gravity by constructing networks whose links are codimension-2 extremal-surfaces  and by  explicitly computing semiclassical commutators of the associated link-areas. Since $d=3,$ codimension-2 extremal-surfaces are geodesics, and codimension-2 `areas' are lengths.      We find a simple 4-link network defined by an HRT surface and a Chen-Dong-Lewkowycz-Qi constrained HRT surface for which all link-areas commute. However, the algebra generated by the link-areas of more general networks tends to be non-Abelian.  One such non-Abelian example is associated with entanglement-wedge cross sections and may be of more general interest.}

\begin{document}
\maketitle
\flushbottom

%%%%%%%%%%%%%%%%%%%%%%%%%%%%%%%%%%%%%%%%%%%%%%%%%%%%%%%%%%%%%%%%%%%%%%%%%%%%%%%%%%%%%%%%%%%%%%%%%%%%%%%%%%%%%%%%%%%%%%%%%%%%%%%%%%%%%%%%%%%%%%%%%%%
\section{Introduction}
Over the last decade, tensor networks have played a key role in developing our understanding of the AdS/CFT correspondence \cite{Maldacena_1999}. They were first proposed as toy models of AdS/CFT in \cite{Swingle_2009, Swingle_2012}, based in part on the observation that the entanglement entropy of a boundary subregion is bounded by an area law that agrees with the Ryu-Takayanagi (RT) formula \cite{Ryu:2006bv, Ryu:2006ef}. It was then shown that certain tensor network constructions saturate this bound \cite{Pastawski_2015, Hayden_2016}. Tensor networks can also model other important aspects of AdS/CFT, including quantum error correction properties \cite{Almheiri:2014lwa} of the holographic dictionary; see e.g. models in \cite{Pastawski_2015, Ferris_2014, Kohler_2019}.

The random tensor networks of \cite{Hayden_2016} have been of particular interest. However, their qualitative properties differ from those of familiar semiclassical bulk states of AdS/CFT as the entanglement spectrum is flat for any boundary region $R$.  By this we mean that the Renyi entropies $S_n$ are approximately independent of $n$.  The same feature arises in the HaPPY code \cite{Pastawski_2015}.   

In the AdS/CFT context, for a given boundary region $R$, and as described in \cite{Akers:2018fow, Dong:2018seb}, producing a state with flat entanglement spectrum requires suppressing fluctuations in the area of the associated Hubeny-Rangamani-Takayanagi (HRT) surface\footnote{I.e., for the covariant generaliztaion of the RT surface.} \cite{Hubeny:2007xt} relative to those in standard semiclassical states.   Bulk states with such suppressed fluctuations are known as \textit{fixed-area states}. 

For a given HRT-surface (associated with a given boundary region $R$), fixed-area states can be produced by projecting more general states onto appropriately-sized windows of HRT-area eigenvalues, perhaps with the window width scaling as $G^{1/2+\epsilon}$ for some small $\epsilon >0$ in terms of the bulk Newton constant $G$.  However, given a set of regions $R_i$, the corresponding collection of entanglement spectra can be rendered flat only if we simultaneously suppress area fluctuations for all of the relevant HRT-surfaces $\gamma_i$.  This in turn requires the associated HRT-area operators to approximately commute.

Unfortunately, as emphasized in \cite{Bao:2018pvs}, commutators of HRT-areas can be large even when all regions $R_i$ lie in a single Cauchy surface of the asymptotically-AdS boundary.  This is in part because the HRT-surfaces $\gamma_i$ generally fail to lie in a single Cauchy surface of the bulk; i.e.,  points on $\gamma_i$ can be causally separated
from points on $\gamma_j$.  The mixing of operators under time-evolution then makes it difficult to avoid sizeable commutators\footnote{In a time-symmetric context, the {\it expectation values} of HRT-area commutators generally vanish.  But the commutators still do not vanish as operators, even if their properties are non-trivial to compute in the semiclassical approximation.}.

One way to address this issue is to modify the notion of a tensor network model following e.g. \cite{Donnelly:2016qqt,Dong:2023kyr}.
However, it is also natural to ask whether the issue can be ameliorated by using the collection of regions $R_i$ to construct a network of HRT-like surfaces that do in fact always lie in a single bulk Cauchy surface, and which thus might potentially have area operators that commute.  Here the use of the term `network' reminds us that a collection of codimension-2 surfaces lying in a (codimension-1) Cauchy surface will generally intersect. One might in particular hope such a network to be related to the tensor network constructions of \cite{Bao:2018pvs,Bao:2019fpq}; see e.g. figure \ref{fig:RT-geo-bao} below.
We emphasize that both the precise notion of what is meant by an HRT-like surface and the extent to which they are useful in producing flat entanglement spectra or the networks of \cite{Bao:2018pvs,Bao:2019fpq} remain to be investigated.  

The present work addresses the first of these steps by considering various constructions of such networks in semiclassical bulk geometries and computing commutators of the areas of the HRT-like surfaces comprising these networks.  We will require our ``HRT-like surfaces"  to be extremal away from points where they intersect other surfaces in the network.  The work below is exploratory, and our goal is merely to investigate a few such networks and collect results that may inform future constructions. 

\begin{figure}[t]
    \centering
     
\begin{tikzpicture}
 
\draw (1,0) arc[start angle=0,delta angle=360,x radius=1cm, y radius=1cm];
\draw [blue] plot [smooth, tension=1] coordinates { (.5,.866) (.25,0) (.5,-.866)};
\draw [blue] plot [smooth, tension=1] coordinates { (0,-1) (.25,-.25) (1,0)};
\draw[blue] (0,1) -- (0,-1);
\draw [blue] plot [smooth, tension=1] coordinates { (-.866,.5) (0,.25) (.866,.5)};
\draw[red] (-.3,.6) -- (-.5,-.2);
\draw[red] (.2,.6) -- (.6,.55);
\draw[red] (-.5,-.2) -- (.13,.03);
\draw[red] (-.3,.6) -- (.2,.6);
\draw[red] (.13,.03) -- (.2,.6);
\draw[red] (.13,.03) -- (.6,.15);
\draw[red] (.6,.15) -- (.6,.55);
\draw[red] (.6,.15) -- (.55,-.4);
\draw[red] (.55,-.4) -- (.2,-.7);
\draw[red] (.2,-.7) -- (.13,.03);

\draw[red] (-.3,.6) -- (-.7,1);
\draw[red] (.2,.6) -- (.45,1.1);
\draw[red] (.6,.55) -- (.9,.85);
\draw[red] (.6,.15) -- (1.1,.3);
\draw[red] (.55,-.4) -- (.9,-.8);
\draw[red] (.2,-.7) -- (.2,-1.2);
\draw[red] (-.5,-.2) -- (-1,-.6);

\fill (-.3,.6) circle (1.5pt);
\fill (.2,.6) circle (1.5pt);
\fill (.6,.55) circle (1.5pt);
\fill (.6,.15) circle (1.5pt);
\fill (.55,-.4) circle (1.5pt);
\fill (.2,-.7) circle (1.5pt);
\fill (-.5,-.2) circle (1.5pt);
\fill (.13,.03) circle (1.5pt);

\end{tikzpicture}
 
    \caption{An example area network and its corresponding tensor network, modelled off of the networks in \cite{Bao:2018pvs}. The area network is shown in blue. The tensors are black nodes, and tensor index contractions are shown as red edges. See \cite{Bao:2018pvs} for explanations.}
    \label{fig:RT-geo-bao}
\end{figure}
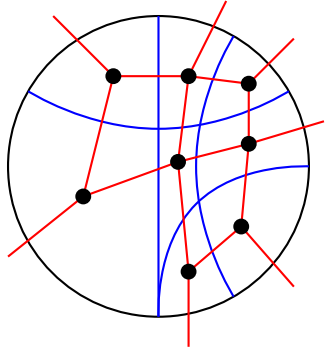

We will analyze the area-operators associated with our network in the semiclassical approximation.  In this context, the operators are described by observables on the classical phase space and their commutators become $i$ times Poisson brackets.  It is then interesting to study the flow generated by such an operator on the classical phase space. Throughout this work we will use the terms ``operator" and ``observable" interchangeably. For HRT-area operators, studies in this direction include \cite{Jafferis:2014lza, Ceyhan:2018zfg, Faulkner:2018faa, Bousso:2019dxk, Bousso:2020yxi, Kaplan_2022}. Much of this work made use of the JLMS formula \cite{Jafferis:2015del} relating the HRT area to the boundary modular Hamiltonian, though see \cite{Kaplan_2022} for a self-contained bulk analysis.  The phase space flow generated by an HRT-area in Einstein-Hilbert gravity turns out to take a simple geometric form that acts as a boundary-condition-preserving kink transformation  (see \cite{Kaplan_2022} for refinements of the discussion in \cite{Bousso:2019dxk, Bousso:2020yxi}).  Extensions to topologically-massive gravity in AdS$_3$ were studied in \cite{Kaplan:2023oaj}. While studies of geometric flow can be of great use, the present work will simply focus on computing commutators associated with our networks and will save analysis of geometric features for future work\footnote{Since we consider areas of surfaces with boundaries, the flow generated by these areas may have a non-trivial effect extending to the boundary, similar to what is found in \cite{Harlow:2021dfp}.}.

We focus on pure Einstein-Hilbert gravity in AdS$_3$ when the boundary metric is 1+1 Minkowski space. In this context we expect that all operators can be expressed in terms of the boundary stress tensor.  An explicit such expression would then allow us to use the    boundary stress tensor algebra to compute arbitrary commutators. While such explicit expressions are difficult to obtain, at the semiclassical level it suffices to work with {\it implicit} expressions as described in \cite{Kaplan_2022}.  The point here is that since the Poisson Bracket $\{A,B\}$ of observables $A,B$ is defined in terms of derivatives of $A,B$ on the phase space, using the chain rule one can use the stress-tensor algebra to compute $\{A,B\}$ even if one knows only the derivatives of $A,B$ with respect to each component of the stress tensor.  Following \cite{Kaplan_2022}, expressions for such derivatives turn out to be straightforward to construct in the sector of our theory given by acting on Poincar\'e AdS$_3$ or a planar black hole with boundary conformal transformations. 

We will consider only this sector below.   We will also refer to these Poisson bracket calculations as "semiclassical commutators" despite the lack of a factor of $i$. Our main results are as follows:

\begin{itemize}
    \item It is generally difficult to construct arbitrarily fine discretizations of the bulk with commuting areas.
    \item We do, however, find a simple 4-link network (analogous to the extremal surface configuration building the four-tensor network of \cite{Bao:2018pvs}) for which the link-areas all commute.
\end{itemize}

The outline of our paper is as follows. Section \ref{sec:setup} reviews the formalism of \cite{Kaplan_2022} for computing Poisson brackets of observables in the above sector of vacuum AdS$_3$. In Section \ref{sec:halfgeo}, we analyze a 4-link \textit{constrained geodesic} network defined by choosing a single HRT surface and two additional boundary-anchor points.  We then follow \cite{Chen:2018rgz} in adding a second surface defined by extremizing the length of a curve connecting the additional two anchors with the constraint that the curve intersects the above-chosen HRT surface. This is the constrained geodesic.  The resulting network is an analogue of the four-tensor network of \cite{Bao:2018pvs}, and we find that all of its areas commute. Appendix \ref{sec:add_surface_app} then analyzes an extension of this simple network, though we find non-vanishing area commutators.

Since the entanglement wedge cross section (EWCS) has been of particular interest in the recent literature \cite{Umemoto_2018, Nguyen_2018, Dutta_2019, Akers_2020, Akers_2022}, we turn to the study of an associated network in section \ref{sec:ewcs}. Again,  this network has non-vanishing area-commutators.  In particular, the EWCS area fails to commute with other areas in the configuration. We conclude with a brief summary and discussion in section \ref{sec:discussion}.

%%%%%%%%%%%%%%%%%%%%%%%%%%%%%%%%%%%%%%%%%%%%%%%%%%%%%%%%%%%%%%%%%%%%%%%%%%%%%%%%%%%%%%%%%%%%%%%%%%%%%%%%%%%%%%%%%%%%%%%%%%%%%%%%%%%%%%%%%%%%%%%%%%%
\section{Commutators from the boundary stress-energy tensor}\label{sec:setup}
The work below will consider pure 2+1 Einstein-Hilbert gravity with negative cosmological constant, and we will restrict attention to solutions that can be obtained from Poincar\'e AdS$_3$ or an $M>0$ planar black hole by acting with boundary conformal transformations\footnote{As described in e.g. \cite{Chua:2023ios}, the full theory consists of a direct sum of disjoint (superselected) phase spaces, each of which can be generated by acting with boundary conformal transformations on any point in the phase space.  The methods used here should thus also be applicable to more general sectors, where one expects them to yield similar results.}.  Furthermore, since solutions in the above classes are equivalent when their boundary stress tensors agree, we may express all observables in our theory in terms of the boundary stress tensor.

In Poincar\'e AdS$_3$, we take the metric to be
\begin{equation}
\label{eq:PAdS3}
ds^2 = \frac{1}{z^2}(-dt^2+dx^2+dz^2) = \frac{1}{z^2}(-dudv+dz^2),
\end{equation}
where we set $l_{AdS}=1$ and we introduce the  light cone coordinates $u=t-x$ and $v=t+x$. 

All of the solutions we consider can be generated from 
\eqref{eq:PAdS3} by acting with boundary conformal transformations. 
In particular, any such transformation can be described by two functions, $U(u)$ and $V(v)$, such that the boundary metric in the solution of interest takes the form
\begin{equation}\label{eq:metric}
    ds^2_\partial = - dU dV =-e^{2\sigma(u,v)}dudv,
\end{equation}with $\sigma(u,v) = \sigma(u) + \hat \sigma(v)$ and hence
\begin{equation}
\label{eq:uUvV}
    \begin{split}
        dU &= e^{2\sigma (u)}du \\
        dV &= e^{2\hat \sigma (v)}dv.
    \end{split}
\end{equation}

The action of a general finite conformal transformation on the stress-energy tensor of a 1+1 dimensional conformal field theory is well known (see e.g. \cite{DiFrancesco:1997nk}) to give
\begin{equation}\label{eq:T}
    T_{ab}dx^a dx^b = T_{ab}^{\text{original}}dx^a dx^b + \frac{c}{12\pi}\biggr[\partial_U^2\sigma_{U} + (\partial_U \sigma_{U})^2\biggr]dU^2 + \frac{c}{12\pi}\biggr[\partial_V^2\hat \sigma_{V} + (\partial_V \hat \sigma_{V})^2\biggr]dV^2,
\end{equation}
where $c$ is the central charge and with $c=3/2G$ for AdS$_3$ Einstein-Hilbert gravity \cite{Brown:1986nw}.   We will choose $U(u), V(v)$ so that the transformation \eqref{eq:uUvV} maps the (vanishing) boundary stress tensor $T_{ab}^\text{original}=0$ of \eqref{eq:PAdS3} to the boundary stress tensor $T_{ab}$ of the desired solution. We then have
\begin{eqnarray}
\label{eq:Tfromsig}
    T_{UU} =& \frac{c}{12\pi}\left[\partial_U^2\sigma_{U} + (\partial_U \sigma_{U})^2\right] \label{eq:TUU}\\
    T_{VV} =& \frac{c}{12\pi}\left[\partial_V^2\hat \sigma_{V} + (\partial_V \hat \sigma_{V})^2\right] \label{eq:TVV}
\end{eqnarray}

In particular, we will define the functions $u(U)$ and $v(V)$ to be the solutions of \eqref{eq:uUvV} subject to certain boundary conditions. And, for a given $T_{ab}$, $\sigma_U$ and $\hat\sigma_V$ are solutions of \eqref{eq:TUU} and \eqref{eq:TVV}, respectively. We will specify the boundary conditions for \eqref{eq:uUvV} and for \eqref{eq:TUU} and \eqref{eq:TVV} at different locations.  To define boundary conditions for \eqref{eq:TUU} and \eqref{eq:TVV} we choose some $U_0,V_0$ and define 
$\sigma_{U_0}(U), \hat \sigma_{V_0}(V)$ to be the solutions of \eqref{eq:uUvV} that satisfy
\begin{eqnarray}
\label{eq:intU0}
\sigma_{U_0}(U)|_{U=U_0} &=& \partial_U \sigma_{U_0}(U)|_{U=U_0} = 0 \cr
\hat \sigma_{V_0}(V)|_{V=V_0} &=& \partial_V \hat \sigma_{V_0}(V)|_{V=V_0}  = 0.    
\end{eqnarray}
In contrast, to define boundary conditions for \eqref{eq:uUvV} we simply note that $U(u), V(v)$ will be defined on intervals $u \in (-\infty,u_{max})$ and $v \in (-\infty,v_{max})$.  We will take $u_{max}=v_{max}=\infty$ for solutions asymptoting to Poincar\'e AdS$_3$ and $u_{max}=v_{max}=0$ for solutions asymptoting to an $M>0$ planar black hole. We choose our boundary conditions to be
\begin{equation}
    u(U=0)=0,\,\,\,\,v(V=0)=0
\end{equation}
for solutions asymptoting to Poincar\'e AdS$_3$ and
\begin{equation}
    u(U=\infty)=0,\,\,\,\,v(V=\infty)=0
\end{equation}
for solutions asymptoting to an $M>0$ planar black hole. In either case, $u_{U_0}(U)$ and $v_{\hat V_0}(V)$ can be written in the form
\begin{eqnarray}\label{eq:utoU}
    u_{U_0}(U) =& \int_{0}^U dU' e^{-2\sigma_{U_0}(U')} + c_u \\
    \label{eq:vtoV} v_{\hat V_0} (V) =& \int_{0}^V dV' e^{-2\hat\sigma_{V_0}(V')} + c_v,
\end{eqnarray}
where, as a consequence of our choice above, $c_u=c_v=0$ when solutions asymptote to Poincar\'e AdS$_3$, while for solutions asymptoting to an $M>0$ black hole we have
\begin{equation}
    c_u = -\int_{0}^{\infty} dU' e^{-2\sigma_{U_0}(U')},\,\,\,c_v=-\int_{0}^{\infty} dV' e^{-2\hat \sigma_{V_0}(V')}.
\end{equation}

As described above, the objects $\sigma_{U_0}$, $\hat \sigma_{V_0}$ are functionals of $T_{ab}$ determined by solving \eqref{eq:TUU} and \eqref{eq:TVV}.  While a closed form solution is not available, we can differentiate \eqref{eq:TUU} with respect to $\sigma_{U_0}(U)$ to obtain a {\it linear} differential equation for $\frac{\delta \sigma_{U_0}(U)}{\delta T_{UU}{(U')}}$.  That linear equation can then be solved to find
\begin{equation}\label{eq:delsdelT}
    \frac{\delta \sigma_{U_0}(U)}{\delta T_{UU}(U')} = \frac{12\pi}{c}  e^{2\sigma_{U_0}(U')}[u_{U_0}(U)-u_{U_0}(U')][\Theta(U-U')\Theta(U'-U_0)-\Theta(U'-U)\Theta(U_0-U')],
\end{equation}
along with the corresponding result for $\hat \sigma_{V_0}(V)$.
Since commutators between boundary stress tensors are given by the Virasoro algebra, the result \eqref{eq:delsdelT} can be used  to compute the Poisson Bracket algebra of conformal factors $\sigma(U,V)$.  Doing so yields a commutator of the form
\begin{equation}\label{eq:sigsig3}
    \begin{split}
        \{\sigma_{U_0}(U),\sigma_{\tilde{U}_0}(\tilde{U})\} =& \frac{6\pi}{c}\bigg[\Theta(\tilde U-U)\\
        &+2(\sigma_{U_0}'(U)-\sigma_{\tilde U_0}'(U))e^{2\sigma_{\tilde U_0}(U)}[u_{\tilde U_0}(\tilde U)-u_{\tilde U_0}(U)]\Theta(\tilde U-U) \\
        &+u_{\tilde{U}_0}(\tilde{U})f_1(U) + f_2(U) + u_{U_0}(U)g_1(\tilde{U}) + g_2(\tilde{U})\bigg],
    \end{split}
\end{equation}
with an analogous expression for $\{\hat \sigma_{V}(V_0),\hat\sigma_{\tilde V_0}(\tilde V)\}$, and we also note that any $\sigma_{U_0}(U)$ commutes with any $\hat \sigma_{V_0}(V)$. The functions $f_{1,2}(U)$ and $g_{1,2}(\tilde{U})$ can depend on $U_0$ and $\tilde{U}_0$, and can be computed explicitly. However, we will not do so here, as we will soon show that their contribution can be ignored. Now, for any two observables $B$ and $C$, we can compute their semiclassical commutator using Eq.~\eqref{eq:sigsig3}:
\begin{equation}\label{eq:leibniz}
\begin{split}
\{B,C\} =
& \int d^2x_1  d^2x'_1 \frac{\delta B}{\delta T_{ij}(x_1)} \{T_{ij}(x_1),T_{i'j'}(x'_1) \} \frac{\delta C}{\delta T_{i'j'}(x_1')} \\
=& \int d^2x_1 d^2x_2 d^2x'_2 d^2x'_1 \frac{\delta B}{\delta \sigma(x_2)} \frac{\delta \sigma(x_2)}{\delta T_{ij}(x_1)} \{T_{ij}(x_1),T_{i'j'}(x'_1) \} \frac{\delta \sigma(x_2')}{\delta T_{i'j'}(x'_1)} \frac{\delta C}{\delta \sigma(x_2')} \\
=& \int d^2x_2 d^2x_2' \frac{\delta B}{\delta \sigma(x_2)}  \{\sigma(x_2),\sigma(x'_2) \} \frac{\delta C}{\delta \sigma(x'_2)},
\end{split}
\end{equation}
in terms of the functional derivatives of $B$ and $C$ with respect to $\sigma(U,V)$.

The above expression \eqref{eq:sigsig3} is rather cumbersome.  At least some part of this is due to the dependence on the unphysical parameters $U_0, \tilde U_0$ associated with the boundary conditions that define $\sigma_{U_0}$ and $\sigma_{\tilde U_0}$.  But physical observables $B,C$ cannot depend on these parameters, so the dependence on $U_0, \tilde U_0$ must cancel completely when computing \eqref{eq:leibniz}.  This suggests that for physical observables $B,C$ it should suffice to use a simplified version of \eqref{eq:sigsig3} that is manifestly independent of $U_0, \tilde U_0$.

In particular, let us recall that, on the space of solutions we choose to study, any physical observable can be written as a functional of the boundary stress tensor. Comparing the three lines of \eqref{eq:leibniz} then shows that
we will obtain the correct commutator $\{B,C\}$ so long as we include some subset of terms from \eqref{eq:sigsig3} that gives the correct expression for $\{T_{ij}(x_1),T_{i'j'}(x'_1) \}$.

In the simple case where $\tilde U_0=U_0$ and $\tilde V_0=V_0$, it turns out that the following effective commutators suffice for this purpose: 
\begin{eqnarray}\label{eq:effcommU}
    \{\sigma_{U_0}(U),\sigma_{{U}_0}(\tilde{U})\}_{eff} =& \frac{6\pi}{c}\Theta(\tilde U-U), \\
    \{\hat{\sigma}_{V_0}(V),\hat{\sigma}_{{V}_0}(\tilde{V})\}_{eff} =& -\frac{6\pi}{c}\Theta(V-\tilde V),
    \label{eq:effcommV}.
\end{eqnarray}
This is straightforward to verify by simply taking appropriate derivatives of \eqref{eq:effcommU} and using~\eqref{eq:TUU} to compute
\begin{eqnarray}
    \{\sigma_{U_0}(U),T_{UU}(\tilde U)\}_{eff} &=& \frac{c}{12\pi} \frac{\partial^2}{\partial \tilde{U}^2} \{\sigma_{U_0}(U),\sigma_{{U}_0}(\tilde{U})\}_{eff} + \frac{c}{6\pi}\sigma_{{U}_0}'(\tilde{U})\frac{\partial}{\partial \tilde{U}}\{\sigma_{U_0}(U),\sigma_{{U}_0}(\tilde{U})\}_{eff} \cr
    &=&\frac{1}{2}\delta'(\tilde U-U) + \sigma_{{U}_0}'(\tilde{U}) \delta(\tilde U-U),
\end{eqnarray}
and thus
\begin{eqnarray}
\label{eq:TTeff}
    \{T_{UU}(U),T_{UU}(\tilde U)\}_{eff} &=& \frac{c}{12\pi} \frac{\partial^2}{\partial U^2} \{\sigma_{U_0}(U),T_{UU}(\tilde{U})\}_{eff} + \frac{c}{6\pi}\sigma_{U_0}'(U)\frac{\partial}{\partial U}\{\sigma_{U_0}(U),T_{UU}(\tilde{U})\}_{eff} \cr 
    &=&\frac{c}{24\pi}\delta'''(\tilde U-U) + \frac{c}{12\pi}\sigma_{{U}_0}'(\tilde{U}) \delta''(\tilde U-U) \cr
    &-&\frac{c}{12\pi}\sigma_{U_0}'(U)\delta''(\tilde U-U) - \frac{c}{6\pi}\sigma_{U_0}'(U)\sigma_{U_0}'(\tilde{U}) \delta'(\tilde U-U).
\end{eqnarray} 
Recall now the following easily verified identities that hold for any smooth functions $f_1(\tilde U, U)$ and $f_2(\tilde U, U)$ that vanish at $\tilde U=U$:
\begin{eqnarray}
\label{eq:f1id}
f_1 \delta''(\tilde U - U) &=& -2\left(\partial_{\tilde U} f_1\right) \partial_{\tilde U} \delta(\tilde U-U) - \left( \partial_{\tilde U}^2 f_1\right) \delta(\tilde U-U) \\
\label{eq:f2id}
f_2 \delta'(\tilde U - U) &=& -\left( \partial_{\tilde U}f_2\right) \delta(\tilde U - U). 
\end{eqnarray}
Using   \eqref{eq:f1id} with $f_1 = \partial_{\tilde U}\sigma_{U_0}(\tilde U) -\partial_{U}\sigma_{U_0}(U)$ then yields
\begin{eqnarray}
\label{eq:TTeff}
    \{T_{UU}(U),T_{UU}(\tilde U)\}_{eff} 
    &=&\frac{c}{24\pi}\delta'''(\tilde U-U) - \frac{c}{6\pi}\left(\sigma''_{U_0}(\tilde U)\right) \partial_{\tilde U} \delta(\tilde U-U)
     \cr
    &-&\frac{c}{12\pi}\left(\sigma'''_{U_0}(\tilde U)\right)\delta(\tilde U-U) - \frac{c}{6\pi}\sigma_{U_0}'(U)\sigma_{U_0}'(\tilde{U}) \delta'(\tilde U-U) \cr
    &=&\frac{c}{24\pi}\delta'''(\tilde U-U) - 2 T_{UU}(\tilde U)  \partial_{\tilde U} \delta(\tilde U-U)
     \cr
    &-&\frac{c}{12\pi}\left(\sigma'''_{U_0}(\tilde U)\right)\delta(\tilde U-U) + \frac{c}{6\pi}\left( [\sigma_{U_0}'(\tilde U)]^2 - \sigma_{U_0}'(U)\sigma_{U_0}'(\tilde{U}) \right) \delta'(\tilde U-U)    \cr 
    &=&\frac{c}{24\pi}\delta'''(\tilde U-U) - 2 T_{UU}(\tilde U)  \partial_{\tilde U} \delta(\tilde U-U)
     -\frac{c}{12\pi}\left(\sigma'''_{U_0}(\tilde U)\right)\delta(\tilde U-U) \cr
     &-& \frac{c}{6\pi}\left( 2\sigma_{U_0}'(\tilde U) \sigma_{U_0}''(\tilde U) - \sigma_{U_0}''(U)\sigma_{U_0}' (\tilde{U})\right) \delta(\tilde U-U) \cr 
        &=&\frac{c}{24\pi}\delta'''(\tilde U-U) - 2 T_{UU}(\tilde U)  \partial_{\tilde U} \delta(\tilde U-U)
     -T_{UU}'(\tilde U) \delta(\tilde U-U) \cr
     &-& \frac{c}{6\pi}\left( \sigma_{U_0}'(\tilde U) \sigma_{U_0}''(\tilde U) - \sigma_{U_0}''(U)\sigma_{U_0}'(\tilde{U})\right)  \delta(\tilde U-U),
\end{eqnarray} 
where the third step used \eqref{eq:f2id} with $f_2 = [\partial_{\tilde U}\sigma_{U_0}(\tilde U)]^2 - \partial_{U}\sigma_{U_0}(U)\partial_{\tilde U}\sigma_{U_0}(\tilde U)$. Since the final term in \eqref{eq:TTeff} vanishes, we see that  \eqref{eq:TTeff} gives the standard stress tensor algebra as desired.

Using the effective commutators \eqref{eq:effcommU}, \eqref{eq:effcommV}, we may thus write the commutator between areas $A_1$ and $A_2$ in the form
\begin{equation}\label{eq:area_comm}
\begin{split}
    \{A_1,A_2\} =& \frac{6\pi}{c}\int_{-\infty}^{\infty} d\tilde U \frac{\delta A_2}{\delta \sigma_{\tilde U_0}(\tilde U)} \int_{-\infty}^{\infty} dU \frac{\delta A_1}{\delta \sigma_{\tilde U_0}(U)}\Theta(\tilde U-U) \\
    &- \frac{6\pi}{c}\int_{-\infty}^{\infty} d\tilde V \frac{\delta A_2}{\delta \hat \sigma_{\tilde V_0}(\tilde V)} \int_{-\infty}^{\infty} dU \frac{\delta A_1}{\delta \hat \sigma_{\tilde V_0}(V)}\Theta(V-\tilde V).
\end{split}
\end{equation}

%%%%%%%%%%%%%%%%%%%%%%%%%%%%%%%%%%%%%%%%%%%%%%%%%%%%%%%%%%%%%%%%%%%%%%%%%%%%%%%%%%%%%%%%%%%%%%%%%%%%%%%%%%%%%%%%%%%%%%%%%%%%%%%%%%%%%%%%%%%%%%%%%%%
\section{A simple constrained-surface network with vanishing commutators}\label{sec:halfgeo}

\begin{figure}
    \centering
    \begin{tikzpicture}[scale=3.5]
    \draw (-1,-1) -- (-.5,1) -- (-.1,.5) -- (-.6,-1.5) -- (-1,-1);

    \draw[dashed, blue] (-.875,-1.16) -- (-.375,.84);
    \draw[dashed, blue] (-.66,-1.42) -- (-.16,.58);
    \draw[dashed, blue] (-.86,-.445) -- (-.46,-.945);
    \draw[dashed, blue] (-.625,.52) -- (-.225,.02);
    
    \draw[teal, thick] (-.73,-.6) arc[start angle=-90,delta angle=33,x radius=15mm, y radius=3.5mm];
    \draw[cyan, thick] (.08,-.54) arc[start angle=-57,delta angle=129,x radius=15mm, y radius=3.5mm];
    \draw[teal] (-.2,-.51) node {$\gamma_1$};
    \draw[cyan] (.7,-.05) node {$\gamma_2$};
    \fill[teal] (-.73,-.6) circle (.4pt);
    \fill[cyan] (-.28,.09) circle (.4pt);
    \draw (-1.1,.9) node {Planar boundary};
    \draw[->] (-.55,-1.5) -- (-.47,-1.17) node [below right] {$U$};
    \draw[->] (-.65,-1.5) -- (-.85,-1.25) node [below left] {$V$};
    
    \draw[red, thick] (-.9,-.8) arc[start angle=-90,delta angle=80,x radius=10mm, y radius=3mm];
    \fill[red] (-.9,-.8) circle (.4pt);
    \draw[violet, thick] (-.5,-.15) arc[start angle=75,delta angle=-85,x radius=8mm, y radius=3.5mm];
    \fill[violet] (-.5,-.15) circle (.4pt);
    \fill[black] (.0805,-.54) circle (.6pt);
    \draw[red] (0,-.75) node {$\gamma_a^{\#}$};
    \draw[violet] (.12,-.3) node {$\gamma_b^{\#}$};

    \draw[red] (-1.1,-.8) node {\tiny $(U_a,V_a)$};
    \draw[violet] (-.5,-.22) node {\tiny $(U_b,V_b)$};
    \draw[teal] (-.89,-.62) node {\tiny $(U_1,V_1)$};
    \draw[cyan] (-.43,.07) node {\tiny $(U_2,V_2)$};
\end{tikzpicture}
    \caption{A constrained geodesic network, with HRT surface $\gamma=\gamma_1 \cup \gamma_2$, and two additional links $\gamma^\#_a$ and $\gamma^\#_b$ that together form a constrained HRT surface $\gamma^\# = \gamma^\#_a \cup \gamma^\#_b$. The anchor points of $\gamma$ are $(U_1,V_1)$ and $(U_2,V_2)$, while $\gamma^\#_a$ is anchored at $(U_a,V_a)$ and $\gamma^\#_b$ at $(U_b,V_b)$.}
    \label{fig:new_config}
\end{figure}
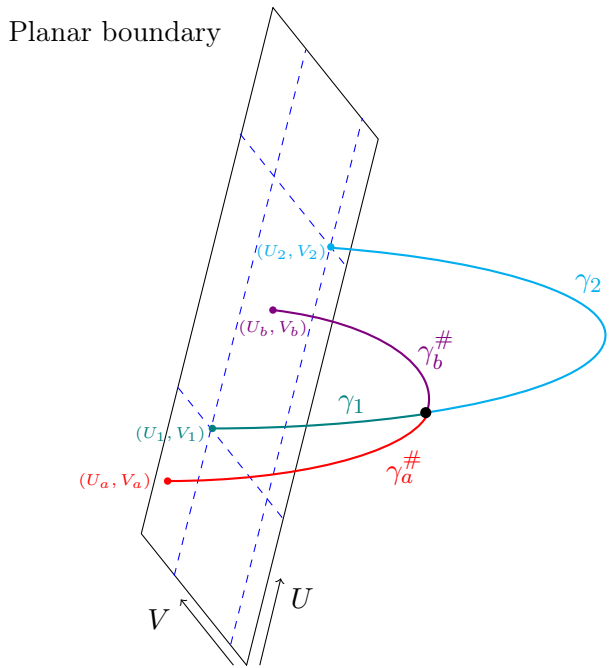

As mentioned in the introduction, we will construct networks of surfaces by extremizing areas subject to constraints that require them to intersect in various ways.  The first such networks will be based on the constrained HRT-surfaces of \cite{Chen:2018rgz}.  Such codimension-2 surfaces $\gamma^\#$ are defined by first choosing an HRT surface $\gamma$ and choosing an anchor set for $\gamma^\#$ on the AdS boundary.  The constrained HRT-surface $\gamma^\#$ is then defined by extremizing its area subject to the usual requirements that its anchors remain fixed and that it satisfy the homology constraint \cite{Headrick:2007km}, but where we also impose the additional constraint that $\gamma^\#$ must intersect $\gamma$; see figure \ref{fig:new_config}. 

The locus of the intersection is then determined by the extremization.  In AdS$_3$, extremal codimension-2 surfaces are geodesics and the intersection occurs at a single point.  In any dimension, the intersection divides $\gamma$ into two half-infinite links $\gamma_1, \gamma_2,$ and it also divides $\gamma^\#$ into $\gamma^\#_a$, $\gamma^\#_b$.   This configuration thus defines a network with a single vertex (at the intersection) and 4 links $\gamma_1, \gamma_2, \gamma^\#_a, \gamma^\#_b$.

Section \ref{sec:halfgeo-area} below computes the renormalized areas of $\gamma^\#_a, \gamma^\#_b$.  
Commutators between the renormalized areas of $\gamma_1, \gamma_2,\gamma^\#_a, \gamma^\#_b$ are then computed in section \ref{sec:halfgeo-results}, where they are shown to vanish.

%%%%%%%%%%%%%%%%%%%%%%%%%%%%%%%%%%%%%%%%%%%%%%%%%%%%%%%%%%%%%%%%%%%%%%%%%%%%%%%%%%
\subsection{Area-operators for half-infinite links}\label{sec:halfgeo-area}

Our task in this section is to compute the areas of $\gamma^\#_a, \gamma^\#_b, \gamma_1, \gamma_2$ for given boundary anchors.  We first focus on $\gamma^\#_a, \gamma^\#_b$.  We take the anchor points of $\gamma$ to be $(U_1,V_1)$ and $(U_2,V_2)$, while $\gamma^\#_a$ is anchored at $(U_a,V_a)$ and $\gamma^\#_b$ at $(U_b,V_b)$.  

It will be convenient to begin with a simple case in Poincar\'e AdS$_3$ where $\gamma$ is in fact defined by the boundary region $R_0$ given by the half-line $x\in [x_1,\infty)$ at some $t=t_1$ on the boundary at $z=0$. Since we are Poincar\'e AdS$_3$, we use the coordinates of \eqref{eq:PAdS3} given by lower-case roman letters. The associated HRT surface $\gamma_{R_0}$ is then just the line of constant $u,v$ with $u=u_1=t_1-x_1$ and $v=v_1=t_1+x_1$ for all $z$. We then define an associated constrained geodesic $\bar{\gamma}^\#$ by choosing two boundary points $(u_a,v_a)$ and $(u_b,v_b)$, where without loss of generality we assume $u_a<u_1<u_b$ and $v_a>v_1>v_b$.   The intersection point then breaks $\bar{\gamma}^\#$ into two half-infinite links $\bar{\gamma}^\#_a, \bar{\gamma}^\#_b$.

\begin{figure}
    \centering
    \begin{tikzpicture}[scale=3]
    \draw (-1,-1) -- (-.5,1) -- (-.1,.5) -- (-.6,-1.5) -- (-1,-1);
    \draw[red, thick] (-.8,-.5) arc[start angle=-90,delta angle=90,x radius=11mm, y radius=2mm];
    \draw[red] (.2,-.5) node {$\bar{\gamma}_a$};
    \fill[red] (-.8,-.5) circle (.4pt);
    \draw[blue, thick] (-.55,-.3) -- (1,-.3);
    \draw[blue] (.8,-.23) node {$\gamma_{R_0}$};
    \fill[blue] (-.55,-.3) circle (.4pt);
    \draw (-1.1,.9) node {Planar boundary};
    \draw[->] (-.55,-1.5) -- (-.47,-1.17) node [below right] {$u$};
    \draw[->] (-.65,-1.5) -- (-.85,-1.25) node [below left] {$v$};
    \draw[dashed, blue] (-.79,-1.26) -- (-.55,-.3) -- (-.29,.74);
    \draw[dashed, blue] (-.75,-.05) -- (-.55,-.3) -- (-.354,-.545);
    \draw[violet, thick] (-.2,.3) arc[start angle=90,delta angle=-90,x radius=5mm, y radius=6mm];
    \fill[violet] (-.2,.3) circle (.4pt);
    \draw[violet] (.1,.28) node {$\bar{\gamma}_b$};
    \fill[black] (.3,-.3) circle (.6pt);
    
\end{tikzpicture}
    \caption{A simple HRT surface $\gamma_{R_0}$ in the vacuum (Poincar\'e AdS$_3$), along with two additional links $\bar{\gamma}^\#_a$ and $\bar{\gamma}^\#_b$ defined by extremizing the area of $\bar{\gamma}^\#_a \cup \bar{\gamma}^\#_b$.}
    \label{fig:original_config}
\end{figure}
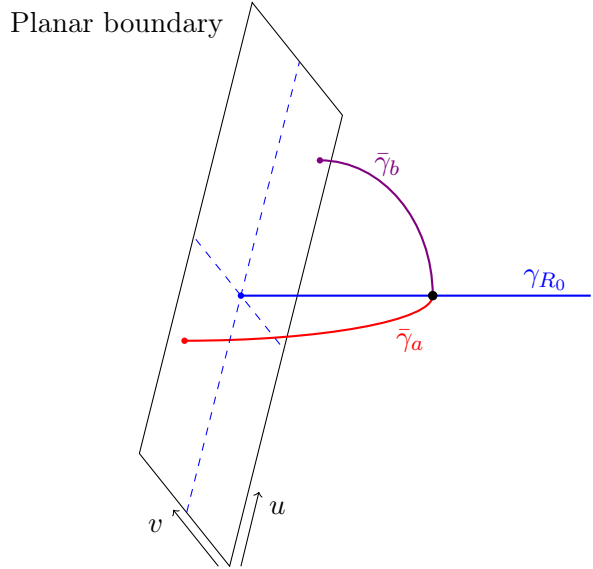

Since the intersection point lies on $\gamma_{R_0}$, it must be of the form $(u_1,v_1,z)$.  But for any half-infinite link $\gamma_{half}$ anchored to $(u_i,v_i)$ on the boundary and the point $(u_1,v_1,z)$ in the bulk, the renormalized area in planar BTZ coordinates with horizon at $z=z_H$ was found in \cite{Morrison_2013} to be 
\begin{equation}
    A_{\gamma_{half}}^{vac} = \ln \biggr( -\frac{2z_H}{z} \biggr[ \sqrt{z_H^2-z^2} \cosh\biggr(\frac{t_1-t_i}{z_H}\biggr) - z_H \cosh\biggr(\frac{x_1-x_i}{z_H}\biggr) \biggr] \biggr).
\end{equation}
In the limit $z_H \to \infty$, the BTZ metric becomes Poincar\'e AdS. Taking this limit, we find the geodesic length
\begin{equation}\label{eq:vac-half}
    A_{\gamma_{half}}^{vac}= \ln \biggr( 2z + \frac{2(u_1-u_i)(v_i-v_1)}{z}  \biggr).
\end{equation}

We now take $\bar{\gamma}^{\#}_a$ to be the half-infinite link with boundary anchor $(u_a,v_a)$, and $\bar{\gamma}^{\#}_b$ to be the half-infinite link with boundary anchor $(u_b,v_b)$. We wish to extremize the total length  $A^{vac}_{\bar{\gamma}^\#_a} + A^{vac}_{\bar{\gamma}^\#_b}$ of $\bar \gamma^\#$ over possible intersection points on $\gamma_{R_0}$.  Since the points on $\gamma_{R_0}$ are labelled by the value of $z$ in \eqref{eq:vac-half}, a short computation yields:
\begin{equation}
\begin{split}
    z_{ext}=[(u_1-u_a)(v_a-v_1)(u_b-u_1)(v_1-v_b)]^{1/4}.
\end{split}
\end{equation}
Inserting this result into \eqref{eq:vac-half} gives
\begin{eqnarray}
    A^{vac}_{\bar{\gamma}^\#_a}=& \ln \bigg[2\bigg(\frac{(u_1-u_a)(v_a-v_1)}{(u_b-u_1)(v_1-v_b)} \bigg)^{1/4}(\sqrt{(u_1-u_a)(v_a-v_1)}+ \sqrt{(u_b-u_1)(v_1-v_b)})
    \bigg]\\
    A^{vac}_{\bar{\gamma}^\#_b}=& \ln \bigg[2\bigg(\frac{(u_b-u_1)(v_1-v_b)}{(u_1-u_a)(v_a-v_1)} \bigg)^{1/4}(\sqrt{(u_1-u_a)(v_a-v_1)}+ \sqrt{(u_b-u_1)(v_1-v_b)})
    \bigg].
\end{eqnarray}

We will now use the above above results to compute similar areas for the general configuration shown in figure \ref{fig:new_config}.  As usual, the idea is to apply an appropriate boundary conformal transformation as in \eqref{eq:metric}. This transformation generates a non-trivial boundary stress tensor, and in that sense takes us out of the vacuum state.  For any half-infinite link area it yields
\begin{eqnarray}\label{eq:Atransform}
    A_{\gamma_{half}} = A^{vac}_{\gamma_{half}} + \sigma_{U_0}(U_i) + \hat{\sigma}_{V_0}(V_i),
\end{eqnarray}
where $U_i=U(u_i)$ and $V_i=V(v_i)$. 

Note that $A^{vac}_{\gamma_{half}}$ depends on the vacuum coordinates $u_i,v_i$ of all three anchor points. Since we wish to fix the physical coordinates $U_i,V_i$ of the anchors, we should regard $u_i,v_i$ as functions of $U_i,V_i$ that depend on some $\sigma_{U_0},\hat{\sigma}_{V_0}$ via \eqref{eq:utoU} and \eqref{eq:vtoV}. Thus, all three terms in Eq.~\eqref{eq:Atransform} can contribute to our commutators.

The last generalization we will need is to transform $\gamma_{R_0}$ into a general HRT-surface $\gamma$ anchored at arbitrary spacelike-separated boundary points $(U_1,V_1)$ and $(U_2,V_2)$. This will also move the other links, transforming our $\bar{\gamma}^\#_a$ to some $\gamma^\#_a$ and taking our $\bar{\gamma}^\#_b$ to some $\gamma^\#_b$. See Figure \ref{fig:new_config}, which shows the result of this transformation.  Without loss of generality, we take $U_a<U_1<U_b<U_2$ and $V_a>V_1>V_b>V_2$. We perform this generalization by taking a fractional linear transformation which brings the second anchor point of $\gamma_{R_0}$ back from infinity, i.e. we take $u \to \frac{1}{u_2-u}$ and $v \to \frac{1}{v_2-v}$. Under this transformation one finds
\begin{equation}
    \sigma_{U_0}(U(u_i)) \to \sigma_{U_0}(U(u_i))+\ln(u_2-u_i),
\end{equation}
with analogous results for $\hat{\sigma}_{V_0}(V)$. As a result, in the general configuration given in Figure \ref{fig:new_config} the link areas take the form
\begin{equation}\label{eq:Agammaa}
\begin{split}
    A_{\gamma^\#_a}=& \ln \bigg[\sqrt{\frac{(u_1-u_a)(v_a-v_1)}{(u_2-u_a)(v_a-v_2)}}+ \sqrt{\frac{(u_b-u_1)(v_1-v_b)}{(u_2-u_b)(v_b-v_2)}}
    \bigg] \\
    &+ \frac{1}{4}\ln\bigg[\frac{(u_1-u_a)(v_a-v_1)(u_2-u_b)(v_b-v_2)(u_2-u_a)^3(v_a-v_2)^3}{(u_2-u_1)^2(v_1-v_2)^2(u_b-u_1)(v_1-v_b)}\bigg] \\
    &+\sigma_{U_0}(U_a) + \hat{\sigma}_{V_0}(V_a) +\ln 2,
\end{split}
\end{equation}
and
\begin{equation}\label{eq:Agammab}
\begin{split}
    A_{\gamma^\#_b}=& \ln \bigg[\sqrt{\frac{(u_1-u_a)(v_a-v_1)}{(u_2-u_a)(v_a-v_2)}}+ \sqrt{\frac{(u_b-u_1)(v_1-v_b)}{(u_2-u_b)(v_b-v_2)}}
    \bigg] \\
    &+ \frac{1}{4}\ln\bigg[\frac{(u_b-u_1)(v_1-v_b)(u_2-u_b)^3(v_b-v_2)^3(u_2-u_a)(v_a-v_2)}{(u_2-u_1)^2(v_1-v_2)^2(u_1-u_a)(v_a-v_1)}\bigg] \\
    &+\sigma_{U_0}(U_b) + \hat{\sigma}_{V_0}(V_b) +\ln 2.
\end{split}
\end{equation}

Our ultimate goal in this calculation is to understand commutators between the areas of the four links  $\gamma^\#_a$, $\gamma^\#_b$, $\gamma_1$, and $\gamma_2$.  Here $\gamma_1$ runs from $(U_1,V_1)$ on the boundary to the intersection point in the bulk, and $\gamma_2$ runs from  $(U_2,V_2)$ on the boundary to the intersection point in the bulk.  It thus remains to compute the areas of $\gamma_1$ and $\gamma_2$ by first calculating the renormalized areas of each piece of $\gamma_{R_0}$, performing the fractional linear transformation $u \to \frac{1}{u_2-u}$ and $v \to \frac{1}{v_2-v}$ to find $A_{\gamma_1}$ and $A_{\gamma_2}$, and finally applying the above conformal transformation. Doing so yields the renormalized areas
\begin{equation}\label{eq:Agamma1}
\begin{split}
    A_{\gamma_1}=& \frac{1}{4}\ln \bigg( \frac{(u_1-u_a)(v_a-v_1)(u_b-u_1)(v_1-v_b)(u_2-u_1)^2(v_1-v_2)^2}{(u_2-u_a)(v_a-v_2)(u_2-u_b)(v_b-v_2)} \bigg) \\
    &+ \sigma_{U_0}(U_1) + \hat{\sigma}_{V_0}(V_1) + \ln 2,
\end{split}
\end{equation}
and
\begin{equation}\label{eq:Agamma2}
    \begin{split}
        A_{\gamma_2}=& \frac{1}{4}\ln \bigg( \frac{(u_2-u_a)(v_a-v_2)(u_2-u_b)(v_b-v_2)(u_2-u_1)^2(v_1-v_2)^2}{(u_1-u_a)(v_a-v_1)(u_b-u_1)(v_1-v_b)} \bigg) \\
    &+ \sigma_{U_0}(U_2) + \hat{\sigma}_{V_0}(V_2) + 2 \ln 2.
    \end{split}
\end{equation}
As a check, adding the above two results one finds the renormalized area of the full geodesic $\gamma=\gamma_1 \cup \gamma_2$ to be
\begin{equation}\label{eq:Agamma}
    A_{\gamma}=\ln[2(u_1-u_a)(v_a-v_1)] + \sigma_{U_0}(U_1) + \hat{\sigma}_{V_0}(V_1) + \sigma_{U_0}(U_2) + \hat{\sigma}_{V_0}(V_2) + 2\ln 2,
\end{equation}
which agrees with \cite{Kaplan_2022}.
As another check, although the above area expressions are written in terms of $\sigma_{U_0}(U)$ and $\hat{\sigma}_{V_0}(V)$, a short computation shows that derivatives of these areas with respect to both $U_0$ or $V_0$ give zero. 
 This is the correct result since $A_{\gamma^\#_a}$, $A_{\gamma^\#_b}$, $A_{\gamma_1}$ and $A_{\gamma_2}$ are physical observables whose definitions do not depend on our arbitrary choice of $U_0$, $V_0$.

In order to calculate commutators, one must take care to express $u,v$ as $\sigma$-dependent functions of $U,V$. After doing so, one may compute the relevant functional derivatives for use in \eqref{eq:leibniz}:
\begin{equation}\label{eq:delAgammaa}
\begin{aligned}
    \frac{\delta A_{\gamma^\#_a}}{\delta \sigma_{U_0}(U)} =& \delta(U-U_a) + e^{-2\sigma_{U_0}(U)}\bigg[-\frac{1+2C}{2(u_1-u_a)}\Theta(U_1-U)\Theta(U-U_a) \\ 
        &-\frac{3-2C}{2(u_2-u_a)}\Theta(U_2-U)\Theta(U-U_a) -\frac{1-2C}{2(u_b-u_1)}\Theta(U_b-U)\Theta(U-U_1) \\ 
        &+\frac{1-2C}{2(u_2-u_b)}\Theta(U_2-U)\Theta(U-U_b) +\frac{1}{u_2-u_1}\Theta(U_2-U)\Theta(U-U_1) \bigg],
\end{aligned}
\end{equation}
\begin{equation}
    \begin{aligned}
        \frac{\delta A_{\gamma^\#_b}}{\delta \sigma_{U_0}(U)} =&      \delta(U-U_b) +  e^{-2\sigma_{U_0}(U)}\bigg[\frac{1-2C}{2(u_1-u_a)}\Theta(U_1-U)\Theta(U-U_a) \\
        &-\frac{1-2C}{2(u_2-u_a)}\Theta(U_2-U)\Theta(U-U_a)-\frac{3-2C}{2(u_b-u_1)}\Theta(U_b-U)\Theta(U-U_1) \\
        &-\frac{1+2C}{2(u_2-u_b)}\Theta(U_2-U)\Theta(U-U_b)+\frac{1}{u_2-u_1}\Theta(U_2-U)\Theta(U-U_1) \bigg]
    \end{aligned}
\end{equation}
\begin{equation}\label{eq:delAgamma1}
    \begin{aligned}
        \frac{\delta A_{\gamma_1}}{\delta \sigma_{U_0}(U)} =&  \delta(U-U_1) + e^{-2\sigma_{U_0}(U)}\bigg[-\frac{1}{2(u_1-u_a)}\Theta(U_1-U)\Theta(U-U_a) \\
        &+\frac{1}{2(u_2-u_a)}\Theta(U_2-U)\Theta(U-U_a) -\frac{1}{2(u_b-u_1)}\Theta(U_b-U)\Theta(U-U_1) \\
        &+\frac{1}{2(u_2-u_b)}\Theta(U_2-U)\Theta(U-U_b) -\frac{1}{u_2-u_1}\Theta(U_2-U)\Theta(U-U_1) \bigg],
    \end{aligned}
\end{equation}
\begin{equation}
    \begin{aligned}
        \frac{\delta A_{\gamma_2}}{\delta \sigma_{U_0}(U)} =&  \delta(U-U_2) + e^{2\sigma_{U_0}(U)}\bigg[-\frac{1}{2(u_1-u_a)}\Theta(U_1-U)\Theta(U-U_a) \\
        &-\frac{1}{2(u_2-u_a)}\Theta(U_2-U)\Theta(U-U_a) +\frac{1}{2(u_b-u_1)}\Theta(U_b-U)\Theta(U-U_1) \\
        &-\frac{1}{2(u_2-u_b)}\Theta(U_2-U)\Theta(U-U_b) -\frac{1}{u_2-u_1}\Theta(U_2-U)\Theta(U-U_1) \bigg],
    \end{aligned}
\end{equation}
with analogous expressions for functional derivatives with respect to $\hat{\sigma}_{V_0}(V)$.  In the above we have defined the quantity
\begin{equation}\label{eq:C}
    C=\frac{\sqrt{\frac{(u_1-u_a)(v_a-v_1)}{(u_2-u_a)(v_a-v_2)}}}{\sqrt{\frac{(u_1-u_a)(v_a-v_1)}{(u_2-u_a)(v_a-v_2)}}+ \sqrt{\frac{(u_b-u_1)(v_1-v_b)}{(u_2-u_b)(v_b-v_2)}}}.
\end{equation}

%%%%%%%%%%%%%%%%%%%%%%%%%%%%%%%%%%%%%%%%%%%%%%%%%%%%%%%%%%%%%%%%%%%%%%%%%%%%%%%%%%%%%%%%%%%%%%%%%%
\subsection{Vanishing commutators for the 4-link constrained HRT-surface network}\label{sec:halfgeo-results}
We now use the above results and \eqref{eq:area_comm} to compute the desired commutators. First, as a check on our results above, let us compute $\{A_{\gamma^\#_a},A_{\gamma}\}$ and $\{A_{\gamma^\#_b},A_{\gamma}\}$. Each of these must vanish since the flow generated by $A_{\gamma}$ is known to introduce a relative boost between the entanglement wedges on either side of $\gamma$ but to preserve the geometry of each wedge separately; see e.g. \cite{Kaplan_2022} which builds on \cite{Bousso:2019dxk,Bousso:2020yxi}.  Since $\gamma^\#_a$ and $\gamma^\#_b$ each lie entirely in one of these wedges, the relative boost has no effect on their areas. Thus their area operators must commute with $A_{\gamma}$. Combining the above equations does indeed yield this result.

We next examine commutators between any two of $\gamma^\#_a$, $\gamma^\#_b$, $\gamma_1$, and $\gamma_2$. A priori, we have no argument for the form that these should take. However, direct calculation shows that all terms cancel.  In particular, the $U$-parts alone give a result of the form
\begin{equation}
    c_1 + c_2 C,
\end{equation}
with constants $c_1$ and $c_2$. For example, $\{A_{\gamma^\#_a},A_{\gamma_1}\}_{U-component} = \frac12(1-C)$. The calculation of the $V$-components then follows immediately: The functional derivatives with respect to $V$ are direct analogues of those with respect to $U$, but with the ordering of the anchor points reversed. However, we must also take into account the various signs that arise in comparing the $U$- and $V$-dependent pieces in \eqref{eq:area_comm}.  The changes inside the step-function are just those associated with the above reversal in the order of the anchor points, but there nevertheless remains an overall difference in sign.  The result of the computation of the $V$-parts is thus identical to that for the $U$-parts up to this overall sign.  Since $C$ is invariant under $u\to v$, this means the $V$-parts of the commutator take the form $-c_1-c_2 C$ so that they precisely cancel the contributions from the $U$-parts.  We thus find that the areas of $\gamma^\#_a$, $\gamma^\#_b$, $\gamma_1$, and $\gamma_2$ mutually commute.

This is an intriguing result. One may then wonder whether similar results hold for other simple networks.  We explore a 6-link example in Appendix \ref{sec:add_surface_app} obtained by adding a further constrained HRT-surface to the network above.  However,  in that case we find link-area commutators that fail to vanish.

%%%%%%%%%%%%%%%%%%%%%%%%%%%%%%%%%%%%%%%%%%%%%%%%%%%%%%%%%%%%%%%%%%%%%%%%%%%%%%%%%%%%%%%%%%%%%%%%%%%%%%%%%%%%%%%%%%%%%%%%%%%%%%%%%%%%%%%%%%%%%%%%%%%
\section{Link-area algebras for the cross section network}\label{sec:ewcs}

It is interesting that the link-area commutators vanished for the constrained HRT-surface network of figure \ref{fig:new_config}.  However, following our original motivations requires us to ask whether the same result can hold in a more complicated network.  While there are clearly many options that one can consider, we focus here on a network associated with the entanglement wedge cross section (see figure \ref{fig:ewcs-ts}), for which the resulting link-area operators may be of interest in their own right.  As before, we begin by finding
expressions for the areas of the entanglement wedge cross section and the four half-infinite HRT-surfaces in section \ref{sec:ewcs-area}. We then compute the various area commutators in Section \ref{sec:ewcs-results}. In contrast to the previous section, we find that some of these commutators do not vanish.

%%%%%%%%%%%%%%%%%%%%%%%%%%%%%%%%%%%%%%%%%%%%%%%%%%%%%%%%%%%%%%%%%%%%%%%%%%%%%%%%%%
\subsection{Area operators}\label{sec:ewcs-area}
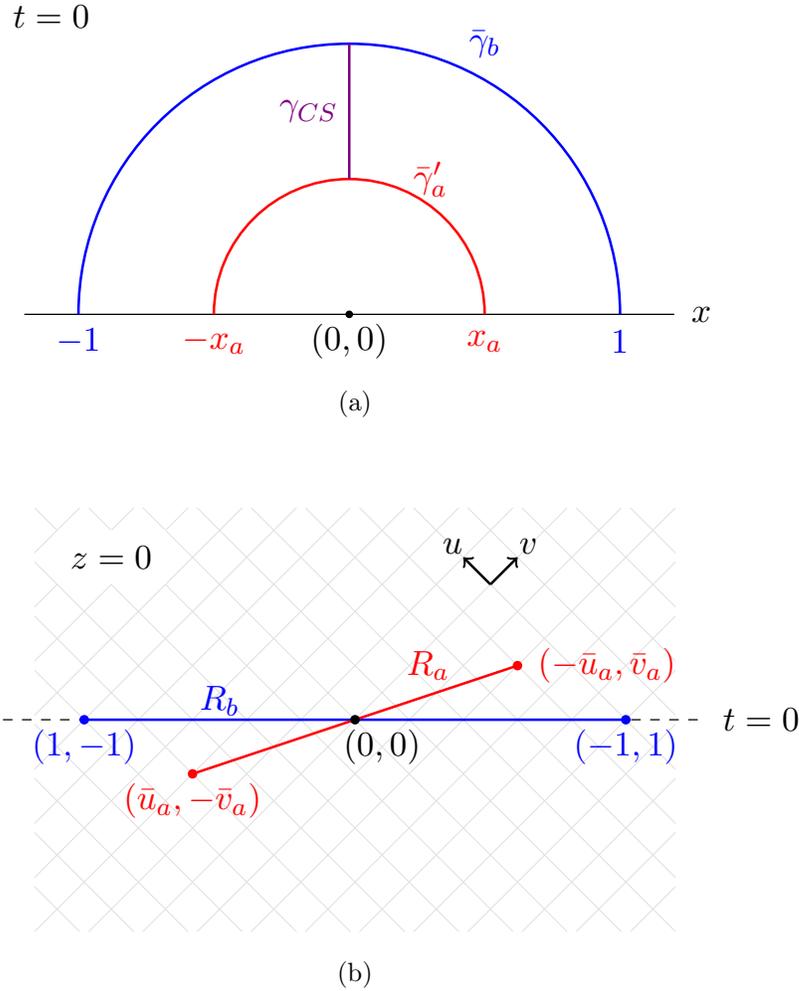
\begin{figure}[t]
    \begin{subfigure}[b]{\textwidth}
    \centering
    \begin{tikzpicture}[scale=3]
    \draw (-1.2,0) -- (1.2,0);
    \draw[blue, thick] (1,0) arc[start angle=0,delta angle=180,x radius=10mm, y radius=10mm];
    \draw[red, thick] (.5,0) arc[start angle=0,delta angle=180,x radius=5mm, y radius=5mm];
    \draw[blue] (1,-.1) node {$1$};
    \draw[blue] (-1,-.1) node {$-1$};
    \draw[red] (.5,-.1) node {$x_a$};
    \draw[red] (-.5,-.1) node {$-x_a$};
    \draw (1.3,0) node {$x$};
    \draw[white] (-1.3,0) node {$x$};
    \fill (0,0) circle (.4pt);
    \draw (0,-.1) node {$(0,0)$};
    \draw (-1.1,1.1) node {$t=0$};
    \draw[violet,thick] (0,1) -- (0,.5);
    \draw[violet] (-.15,.75) node {$\gamma_{CS}$};
    \draw[blue] (.5,1) node {$\bar{\gamma}_b$};
    \draw[red] (.3,.5) node {$\bar{\gamma}_a'$};
\end{tikzpicture}
    \caption{}
    \label{fig:ewcs-ts}
    \vspace{1cm}
  \end{subfigure}
  \begin{subfigure}[b]{\textwidth}
    \centering
    \begin{tikzpicture}[scale=3]

    \def\a{1.2}
    \def\b{.8}
    \def\lw{0.1}
    \draw[] (-\a,-\b) rectangle (\a,\b);
    \foreach \x in{0,0.2,0.4,...,2.4}{
    \draw[gray!25,line width=\lw mm](-\a+\x,-\b)--({less(\a,2*\b+\x-\a)==1?\a:(2*\b+\x-\a)}, {less(2*\a-\b-\x,\b)==1?(2*\a-\b-\x):\b});
    \draw[gray!25,line width=\lw mm](-\a+\x,\b)--({less(\a,2*\b+\x-\a)==1?\a:(2*\b+\x-\a)}, {less(2*\a-\b-\x,\b)==1?(-2*\a+\b+\x):(-\b)});
    }
    \foreach \x in{0.2,0.4,...,1.6}{
    \draw[gray!25,line width=\lw mm](-\a,-\b+\x)--(2*\b-\x-\a, \b);
    \draw[gray!25,line width=\lw mm](-\a,\b-\x)--(2*\b-\x-\a, -\b);
    }

    \draw [white,line width = 1 mm] (-1.2,-.8) rectangle (1.2,.8);
    \draw (1.5,0) node {$t=0$};
    \draw[white] (-1.5,0) node {$t=0$};
    \fill[white] (-1.1,.5) rectangle (-.7,.7);
    \draw (-.9,.6) node {$z=0$};
    \draw[dashed]  (-1.3,0) -- (1.3,0);

    \draw[->,thick] (.5,.5) -- (.4,.6);
    \draw[->,thick] (.5,.5) -- (.6,.6);
    \draw (.36,.64) node {$u$};
    \draw (.64,.64) node {$v$};
    
    \draw[blue] (-1,-.1) node {$(1,-1)$};
    \fill[blue] (-1,0) circle (.5pt);
    \draw[blue] (1,-.1) node {$(-1,1)$};
    \fill[blue] (1,0) circle (.5pt);
    \draw[blue,thick] (-1,0) -- (1,0);
    
    \draw[red,thick] (-.6,-.2) --  (.6,.2);
    \draw[red] (-.6,-.3) node {$(\bar u_a,-\bar v_a)$};
    \fill[red] (-.6,-.2) circle (.5pt);
    \draw[red] (.93,.2) node {$(-\bar u_a,\bar v_a)$};
    \fill[red] (.6,.2) circle (.5pt);
    \draw (.1,-.1) node {$(0,0)$};
    \fill (0,0) circle (.5pt);

    \draw[blue] (-.5,.07) node {$R_b$};
    \draw[red] (.27,.2) node {$R_a$};

\end{tikzpicture}
    \caption{}
    \label{fig:ewcs-boost}
  \end{subfigure}
  \caption{A simple example of a cross-section network. Panel (a) shows the $t=0$ slice, which we take to contain two HRT surfaces $\bar{\gamma}_a'$ and $\bar{\gamma}_b$. Although the figure shows $x_a$ less than $1$, any value $x_a>0$ is allowed. Panel (b) shows the $z=0$ boundary for a more general configuration in which $R_a$ and $R_b$ are boundary regions respectively homologous to $\bar{\gamma}_a$ and $\bar{\gamma}_b$, but with $R_a$ (and thus also $\bar{\gamma}_a$) now boosted relative to
  $R_b$ and $\bar{\gamma}_b$ (which continue to lie in
  the $t=0$ slice).}
\end{figure}

Our goal in this section is to find expressions for the areas of all of the links in the network shown in Figure \ref{fig:ewcs-ts}, but in the context of a general spacetime in our phase space (i.e., with a general $T_{ab}$ in the allowed class) and with general positions of the anchor points. We consider in detail only cases where the cross section $\gamma_{CS}$ is spacelike (though we briefly comment on the case when the cross section is timelike in Appendix \ref{sec:timelikeCS-app}). The network is defined by first choosing two HRT surfaces, $\bar{\gamma}_a'$ and $\bar{\gamma}_b$, and constructing the associated cross-section $\gamma_{CS}$, defined as the codimension-2 surface whose boundaries lie on the above HRT-surfaces and which has extremal length\footnote{As usual, if there is more than one such surface we would choose the minimal one.  We should also enforce a homology constraint.  However, neither of these details are relevant in the simple context studied here.}.  In particular, this extremization condition fixes the locations of the cross-section boundaries on the original HRT-surfaces.  When the region between the HRT-surfaces is an entanglement wedge, this construction defines an entanglement-wedge cross section (though our computation holds more generally).

As before, we will generate general configurations by acting on simple ones with boundary conformal transformations. We start in the Poincar\'e AdS$_3$ vacuum and choose two boundary subregions, $R_a$ and $R_b$.  Both regions are to be defined by straight line segments on the boundary, though they need not lie in any $t=constant$ slice.  However, we can simplify the configuration by acting with boundary conformal transformations that act on the boundary as fractional linear transformations in either $u$ or $v$, as such boundary conformal transformations preserve the vanishing of the boundary stress tensor.  The resulting 6-parameter group can generally\footnote{The only exceptions correspond to cases where $R_a$ and $R_b$ define a common Lorentz frame  but fail to share a common time-slice.  Such exceptions can be treated as degenerate limits of the more general case studied explicitly.} be used to move both $R_a$ and $R_b$ to line segments that are symmetric about the origin  $(u,v)=(0,0)$ of the boundary Minkowski space; see figure \ref{fig:ewcs-boost}. 

These conditions fix a 4-parameter subgroup of the above symmetries, but they still allow further action by both boosts and dilations.  It is convenient to use the boosts to place the segment $R_b$ in the surface $t=0$. 

Indeed, it will be useful to begin with an even simpler class of configurations in which all anchor points lie in a constant time slice as shown in Figure \ref{fig:ewcs-ts}.   We emphasize that this configuration is no longer related by symmetries to the most general ones, but we will see that it is nevertheless useful starting point for our analysis.  We choose the anchor points of the first HRT surface, $\bar{\gamma}_a'$, to lie at $x=\pm x_a$, with $x_a>0$. The anchor points of the second HRT surface, $\bar{\gamma}_b$, are fixed at $x = \pm 1$. Given any 4 values of $x$ one can define a useful cross-ratio (see also \eqref{eq:chiuv2} below) which for these  anchor points takes the value
\begin{equation}\label{eq:chi-ts}
    \chi = \frac{4x_a}{(x_a+1)^2}.
\end{equation}

From this we can compute the area of the cross section, either directly or by using results from \cite{Caputa_2019,Dutta_2019}. We find
\begin{equation}\label{eq:AEWCS-ts}
    A_{CS} = \ln \bigg( \frac{1+\sqrt{1-\chi}}{\sqrt{\chi}}\bigg) = \frac{1}{2} |\ln x_a|.
\end{equation}
Notice that, since the cross-section $\gamma_{CS}$ does not extend to the AdS boundary, its renormalized area is just its (finite) area.  As a result, for any solution in our phase space (with general $T_{ab}$) the cross-section area $A_{CS}$ continues to be given by \eqref{eq:AEWCS-ts} so long as $u,v$ are expressed in terms of the physical coordinates $U,V$.

We would now like to generalize the configuration in Figure \ref{fig:ewcs-ts} by boosting $R_a$, the boundary region that defines $\bar{\gamma}_a'$, relative to $R_b$ as shown in Figure \ref{fig:ewcs-boost}. We will denote the resulting HRT surface by $\bar{\gamma}_a$, with anchor points at $(\bar{u}_a,-\bar{v}_a)$ and $(-\bar{u}_a,\bar{v}_a)$. We take $\bar{u}_a>0$ and $\bar{v}_a>0$. 

Note that the cross-section itself is invariant under this boost. Since the result of any boost satisfies
\begin{equation}
    x_a^2 = \bar u_a \bar v_a,
\end{equation}
we can write \eqref{eq:AEWCS-ts} in terms of the anchor points in this new configuration to find
\begin{equation}\label{eq:AEWCS-symm}
    A_{CS} = \frac{1}{4}|\ln \bar u_a \bar v_a|.
\end{equation}
We can also write down the vacuum values of the areas of the four half-infinite links in our network.  For any HRT surface with anchor points $(\bar{u},-\bar{v})$ and $(-\bar{u},\bar{v})$, the vacuum area of each half-infinite link is given by Eq.~\eqref{eq:vac-half} with $(u_2,v_2)=(0,0)$ and $z=\sqrt{\bar{u}\bar{v}}$. This yields
\begin{equation}
    A^{vac}_{\gamma_{half}} = \frac{1}{2}\ln [4(\bar u-(-\bar u))(\bar v-(-\bar v))],
\end{equation}
where we have made manifest the contributions from each anchor point. Thus, the vacuum areas of the links cut from $\bar{\gamma}_a$ and $\bar{\gamma}_b$ are
\begin{eqnarray}
    A^{vac}_{\bar{\gamma}_{a,half}} = \frac{1}{2}\ln [4(\bar u_a-(-\bar u_a))(\bar v_a-(-\bar v_a))], \label{eq:Aahalf}\\
    A^{vac}_{\bar{\gamma}_{b, half}} = \frac{1}{2}\ln [4(\bar u_b-(-\bar u_b))(\bar v_b-(-\bar v_b))],\label{eq:Abhalf}
\end{eqnarray}
where we have $\bar u_b = 1$ and $\bar v_b = 1$. Writing $A^{vac}_{\bar{\gamma}_{b,half}}$ in the somewhat awkward form above will turn out to clarify later calculations. We can now apply the conformal transformation \eqref{eq:metric}, under which the vacuum areas above transform as in \eqref{eq:Atransform}. This gives the area of the half-infinite links cut from $\bar{\gamma}_a$ and $\bar{\gamma}_b$ in any solution.

We can now move our anchor points into an almost fully general configuration in the Poincar\'e AdS$_3$ vacuum by acting with an  SL(2,$\mathbb{R}$) $\times$ SL(2,$\mathbb{R}$) transformation. We wish to find the transformation that takes two general HRT surfaces, $\gamma_a$ and $\gamma_b$, into the previous configuration, transforming $\gamma_a$ into $\bar{\gamma}_a$ and $\gamma_b$ into $\bar{\gamma}_b$. If $\gamma_a$ has anchor points $( u_{a1},v_{a1})$ and $(u_{a2}, v_{a2})$, then we have the constraints
\begin{equation}\label{eq:constraintsa}
    \frac{a_u u_{a1}+b_u}{c_u u_{a1} + d_u} = \bar u_a,\,\,\,\frac{a_u u_{a2}+b_u}{c_u u_{a2} + d_u} = -\bar u_a,\,\,\,\frac{a_v v_{a1}+b_v}{c_v v_{a1} + d_v}=-\bar v_a,\,\,\,\frac{a_v v_{a2}+b_v}{c_v v_{a2} + d_v}=\bar v_a.
\end{equation}
Additionally, if $\gamma_b$ has anchor points $( u_{b1}, v_{b1})$ and $( u_{b2}, v_{b2})$, then we have
\begin{equation}\label{eq:constraintsb}
\begin{split}
    &\frac{a_u u_{b1}+b_u}{c_u u_{b1} + d_u} = \bar{u}_b = 1,\,\,\,\frac{a_u u_{b2}+b_u}{c_u u_{b2} + d_u} = -\bar{u}_b = -1,\\
    &\frac{a_v v_{b1}+b_v}{c_v v_{b1} + d_v}= -\bar{v}_b = -1,\,\,\,\frac{a_v v_{b2}+b_v}{c_v v_{b2} + d_v}= \bar{v}_b = 1.
\end{split}
\end{equation}
We are also free to impose the additional constraints $a_u=a_v=1$. We can then solve for $\bar{u}_a$ and $\bar{v}_a$ in terms of the four anchor points of our general HRT surfaces. In terms of the cross ratios 
\begin{equation}
\label{eq:chiuv2}
    \chi_u = \frac{(u_{a2}-u_{a1})(u_{b2}-u_{b1})}{(u_{a1}-u_{b2})(u_{b1}-u_{a2})},\,\,\,\,\chi_v = \frac{(v_{a2}-v_{a1})(v_{b2}-v_{b1})}{(v_{a1}-v_{b2})(v_{b1}-v_{a2})},
\end{equation}
we find\footnote{The constraints \eqref{eq:constraintsa} and \eqref{eq:constraintsb} admit two solutions, differing by a sign in front of the square root term in the numerator. We choose the sign consistent with the case where both intervals lie in the $t=0$ surface.}
\begin{equation}
    \bar u_a = \left(\frac{1+\sqrt{1-\chi_u}}{\sqrt{\chi_u}} \right)^2, \,\,\,\, \bar v_a = \left(\frac{1+\sqrt{1-\chi_v}}{\sqrt{\chi_v}} \right)^2.
\end{equation}
Since $\chi_u, \chi_v <1$, the expressions for $\bar{u}_a$, $\bar{v}_a$ are real\footnote{The cross-ratios can be written as $\chi_u= \frac{4\bar{u}_a}{(\bar{u}_a+1)^2}$ and $\chi_v=\frac{4\bar{v}_a}{(\bar{v}_a+1)^2}$, which are less than 1 for any $\bar{u}_a \neq 1$ and $\bar{v}_a \neq 1$, respectively. This can also be argued directly from the form of $\chi_{u,v}$ in Eq.~\eqref{eq:chiuv2}.}. Using these definitions in \eqref{eq:AEWCS-symm} then yields
\begin{equation}
    A_{CS} = \frac{1}{2} \left( \ln \left[\frac{1+\sqrt{1-\chi_u}}{\sqrt{\chi_u}} \right] + \ln \left[\frac{1+\sqrt{1-\chi_v}}{\sqrt{\chi_v}} \right] \right),
\end{equation}
where we have dropped the absolute value sign since the expression is manifestly positive (the arguments in the logarithms are greater than one since $\chi_u, \chi_v <1$). Note that this reduces to the result \eqref{eq:AEWCS-symm} when all anchor points lie on a slice with time-reversal symmetry, since in that case $\chi_u=\chi_v$.

The four half-infinite links are now $\gamma_{a1}$ anchored to $(u_{a1}, v_{a1})$, $\gamma_{a2}$ anchored to $(u_{a2}, v_{a2})$, $\gamma_{b1}$ anchored to $(u_{b1}, v_{b1})$, and $\gamma_{b2}$ anchored to $(u_{b2}, v_{b2})$. Writing the constraints with parameters $b_{u,v}$, $c_{u,v}$, and $d_{u,v}$ expressed in terms of the anchor points, and using the conformally transformed versions of Equations \eqref{eq:Aahalf} and \eqref{eq:Abhalf}, we obtain the final expressions for the areas of the half-infinite links 
\begin{equation}\label{eq:Aa1}
    \begin{split}
        A_{a1}=&\frac{1}{4}\ln\bigg|\frac{(u_{a1}-u_{b1})(u_{a1}-u_{b2})(v_{a1}-v_{b1})(v_{a1}-v_{b2})}{(u_{b1}-u_{a2})(u_{b2}-u_{a2})(v_{b1}-v_{a2})(v_{b2}-v_{a2})}\bigg| \\
        &+\frac{1}{2}\ln|4(u_{a1}-u_{a2})(v_{a2}-v_{a1})| + \sigma_{U_0}( U_{a1})+\sigma_{V_0}(V_{a1}),
    \end{split}
\end{equation}
\begin{equation}
    \begin{split}
        A_{a2}=&\frac{1}{4}\ln\bigg|\frac{(u_{b1}-u_{a2})(u_{b2}-u_{a2})(v_{b1}-v_{a2})(v_{b2}-v_{a2})}{(u_{a1}-u_{b1})(u_{a1}-u_{b2})(v_{a1}-v_{b1})(v_{a1}-v_{b2})}\bigg| \\
        &+\frac{1}{2}\ln|4(u_{a1}-u_{a2})(v_{a2}-v_{a1})| + \sigma_{U_0}( U_{a2})+\sigma_{V_0}(V_{a2}), 
    \end{split}
\end{equation}
\begin{equation}
    \begin{split}
        A_{b1}=&\frac{1}{4}\ln\bigg|\frac{(u_{a1}-u_{b1})(v_{a1}-v_{b1})(u_{b1}-u_{a2})(v_{b1}-v_{a2})}{(u_{a1}-u_{b2})(v_{a1}-v_{b2})(u_{b2}-u_{a2})(v_{b2}-v_{a2})}\bigg| \\
        &+\frac{1}{2}\ln|4(u_{b1}-u_{b2})(v_{b2}-v_{b1})| + \sigma_{U_0}( U_{b1})+\sigma_{V_0}(V_{b1}),
    \end{split}
\end{equation}
\begin{equation}\label{eq:Ab2}
    \begin{split}
        A_{b2}=&\frac{1}{4}\ln\bigg|\frac{(u_{a1}-u_{b2})(v_{a1}-v_{b2})(u_{b2}-u_{a2})(v_{b2}-v_{a2})}{(u_{a1}-u_{b1})(v_{a1}-v_{b1})(u_{b1}-u_{a2})(v_{b1}-v_{a2})}\bigg| \\
        &+\frac{1}{2}\ln|4(u_{b1}-u_{b2})(v_{b2}-v_{b1})| + \sigma_{U_0}( U_{b2})+\sigma_{V_0}(V_{b2}). \\
    \end{split}
\end{equation}

Having found these results, it is also useful to note that, if we had been satisfied with less detailed knowledge of the above functions, we could have obtained certain information about these areas by following a much simpler route.  In particular, since each half-infinite HRT-surface has a single anchor point,  it is manifest that each such area transforms covariantly under boundary conformal transformations, and that it transforms as the logarithm of a local operator that has conformal weight $1$ at its anchor point. 

In particular, these areas must transform in this way under the SL(2,$\mathbb{R}$) $\times$ SL(2,$\mathbb{R}$) group of fractional linear transformations.  Since any SL(2,$\mathbb{R}$) $\times$ SL(2,$\mathbb{R}$)-invariant function of four anchor points is a function of the cross-ratios $\chi_u, \chi_v$, for e.g. $\gamma_{b1}$ we must have 
\begin{equation}\label{eq:Ab1-form}
\begin{split}
    A_{b1} =& \frac{1}{2}\ln\left(\frac{(u_{a1}-u_{b1})(v_{b1}-v_{a1})(u_{a2}-u_{b1})(v_{b1}-v_{a2})}{(u_{a2}-u_{a1})(v_{a1}-v_{a2})}\right) \\
    &+ f_{b1}(\chi_u, \chi_v) + \sigma_{U_0}( U_{b1})+\sigma_{V_0}(V_{b1}), 
\end{split}
\end{equation}
where $f_{b1}(\chi_u,\chi_v)$ is some (separable) function of the cross ratios that can be found by comparing with \eqref{eq:Aa1}.  The  areas of the other three half-infinite links take similar forms. This form turns out to be useful in simplifying some of the commutator calculations since all functions of $u$-coordinates commute with all functions of $v$-coordinates so that we also have $\{\chi_u, f_{b1}(\chi_u,\chi_v)\}=0$.

It is now a straightforward exercise to compute functional derivatives of link areas with respect to $\sigma(U,V)$, after first expressing the area operators in terms of coordinates $U,V$ using \eqref{eq:utoU} and \eqref{eq:vtoV}. We can do this for the half-infinite link areas as well as for the cross-section area. We save detailed expressions for Appendix \ref{sec:func-derivs-app}, but note here that for the area $A_{CS}$ of the cross-section $\gamma_{CS}$, we can write
\begin{equation}\label{eq:delAewcs}
\begin{split}
    \frac{\delta A_{CS}}{\delta \sigma_{U}(U)} =& \frac{\partial A_{CS}}{\partial \chi_u} \frac{\delta \chi_u}{\delta \sigma_{U}(U)} \\
    =& -\frac{1}{4 \chi_u \sqrt{1-\chi_u}} \frac{\delta \chi_u}{\delta \sigma_{U}(U)},
\end{split}
\end{equation}
and similarly for $\frac{\delta A_{CS}}{\delta \hat{\sigma}_{V_0}(V)}$. As in section \ref{sec:halfgeo}, such formulae can be combined with \eqref{eq:leibniz} to compute the desired semiclassical commutators.  We discuss the results in section \ref{sec:ewcs-results} below.

%%%%%%%%%%%%%%%%%%%%%%%%%%%%%%%%%%%%%%%%%%%%%%%%%%%%%%%%%%%%%%%%%%%%%%%%%%%%%%%%%%
\subsection{Results}\label{sec:ewcs-results}
We now compute commutators between (i) any two half-infinite link areas and (ii) the cross-section area and any of the half-infinite link areas. In general, commutators of type (i) vanish, but those of type (ii) are non-zero.

Let us start by computing commutators of type (i). Commutators between half-infinite links on the same HRT surface must vanish since these are equivalent to commutators between a half-infinite link and the HRT surface containing it. HRT area flow leaves the HRT surface invariant, and so it should leave the link area unaffected. However, to understand commutators between half-infinite links on \textit{different} HRT surfaces, we must perform a calculation. We do this by first taking the link-area functional derivatives given in Equations \eqref{eq:delAa1}-\eqref{eq:delAb2}, then using them in Eq.~\eqref{eq:area_comm}. We find that all such commutators vanish. And, unlike the link-area algebra of Section \ref{sec:halfgeo-results}, the $U$- and $V$-components of these commutators vanish individually.

The commutators between $A_{CS}$ and the half-infinite link areas are more interesting. We will focus on $\{A_{CS},A_{b1}\}$. Since $A_{CS}$ depends only on $\chi_{u}$ and $\chi_v$, we can use Eq.~\eqref{eq:Ab1-form} for $A_{b1}$ and ignore the $\chi$-dependent piece $f_{b1}(\chi_u,\chi_v)$, since this commutes with all functions of $\chi_{u}$ and $\chi_v$.  We choose the ordering $U_{b1}<U_{a1}<U_{a2}<U_{b2}$ and $V_{b1}>V_{a1}>V_{a2}>V_{b2}$. Using the functional derivative of $A_{b1}$ in Eq.~\eqref{eq:delAb1-form}, and the functional derivative of $A_{CS}$ in Eq.~\eqref{eq:delACS-full} (and their $V$-dependent counterparts), we find
\begin{equation} 
    \{A_{CS},A_{b1}\} = \frac{3\pi}{2c}\sqrt{1-\chi_u} - \frac{3\pi}{2c}\sqrt{1-\chi_v}.
\end{equation}
A similar non-vanishing result is of course obtained when $b1$ is replaced with any other half-infinite links (the result will be the same up to an overall sign). The relative sign difference between $U$-components and $V$-components appears for the same reason as described in Section \ref{sec:halfgeo-results}. We will hence always have a difference in overall sign between commutators with $\chi_u$ and $\chi_v$; otherwise, they will be the same up to a replacement of all instances of $\chi_u$ with $\chi_v$.

Furthermore, one may check that indeed 
\begin{equation}
    \{A_{CS},A_{\gamma_{b1}\cup \gamma_{b2}}\} = \{A_{CS},A_{\gamma_{b1}}\} +\{ A_{CS},A_{\gamma_{b2}}\} = 0,
\end{equation}
as is required by the fact that $\gamma_{b1}\cup \gamma_{b2}$ is an HRT surface relative to which $\gamma_{CS}$ lies entirely in one of the associated entanglement wedges. In particular, since the action of the HRT area is just to introduce a relative boost between the two entanglement wedges, the area $A_{CS}$ is unaffected. The same result holds for $\gamma_{a1}\cup\gamma_{a2}$. As an aside, we note that these HRT areas will fail to commute with $A_{CS}$ if we allow the cross section to be timelike. We elaborate on this result in Appendix \ref{sec:timelikeCS-app}.

%%%%%%%%%%%%%%%%%%%%%%%%%%%%%%%%%%%%%%%%%%%%%%%%%%%%%%%%%%%%%%%%%%%%%%%%%%%%%%%%%%%%%%%%%%%%%%%%%%%%%%%%%%%%%%%%%%%%%%%%%%%%%%%%%%%%%%%%%%%%%%%%%%%%%%%%%%%%%%%%%%%%
\section{Discussion}\label{sec:discussion}

Motivated by a desire to improve the understanding of tensor-network models of holography, our work above probed the feasibility of simultaneously fixing the areas of all surface segments in an area-network. We studied several such area-networks in the context of pure AdS${}_3$ Einstein-Hilbert gravity (where the areas are in fact lengths and extremal codimension-2 surfaces are geodesics) and computed the relevant commutators at leading order in the semiclassical approximation.  

Our first network contained precisely 4 links and was defined by a single HRT surface and a single constrainted HRT-surface.    All link-area commutators in this network were found to vanish at leading semiclassical order.  While higher-order effects remain to be considered, they would necessarily be small.  We thus conclude that, at least in the pure-AdS${}_3$ context, this network is one for which fluctuations of all areas can be simultaneously suppressed relative to the $O(\sqrt{G})$ fluctuations found in typical semiclassical states.

However, such results are not generic.  In particular, Appendix \ref{sec:add_surface_app} analyzed a 6-link generalization of the above model defined by adding an additional constrained geodesic. For this network we found non-vanishing commutators.

We then moved on to study a network with two HRT surfaces and their cross-section $\gamma_{CS}$.  When the region between the two HRT surfaces is an entanglement wedge, this  $\gamma_{CS}$ in the associated entanglement-wedge cross-section.  Here we again found non-vanishing area commutators. 

The present work was exploratory and did not seek deeper understanding of the results. It is thus far from clear that we have exhausted the space of interesting constructions, though it is also unclear which additional area-networks would be of significant interest for further study.  On the other hand, it {\it would} clearly be of interest to understand whether the area-link algebra found for the 4-link constrained-HRT network of section \ref{sec:halfgeo} remains Abelian in theories with matter and/or in higher dimensions.  If it does, the result would then call out for an explanation or interpretation in terms of a dual CFT.

Another issue to which we expect to return is the question of obtaining a more geometric understanding of the commutators described above and the flows generated by them.  The fact that HRT areas are known to generate flows on phase space described by geometric operations \cite{Bousso:2019dxk, Bousso:2020yxi,Kaplan_2022} that something similar may be true of the HRT area-links studied in the present work. And since such links have boundaries, it is natural to expect the flow generated by these areas may have a non-trivial effect extending to the boundary, as in \cite{Harlow:2021dfp} for similar operators associated with codimension-1 surfaces. Such an understanding might be particularly useful in the context of entanglement-wedge cross-sections, where the cross-section area is associated with reflected entropy \cite{Dutta_2019} and has been related to entanglement of purification \cite{Umemoto_2018,Nguyen_2018}.   These issues will be addressed in forthcoming work.

\acknowledgments
The work of JH, MK, and DM was supported in part by the U.S. Air Force Office of Scientific Research under award number FA9550-19-1-0360. JH also acknowledges support from the U.S. Department of Defence through the National Defense Science and Engineering Graduate (NDSEG) Fellowship Program. J-qW was supported in part by a DeBenedictis postdoctoral Fellowship.  J.q.W is supported by the National Natural Science Foundation of China (NSFC) Project No.12047503.  This work was also supported in part by funds from the University of California.
We thank Pratik Rath for suggestions that led to some of the configurations studied in this work and for his contributions in early stages of the analysis.

\appendix
\section{Adding additional constrained HRT surfaces}\label{sec:add_surface_app}
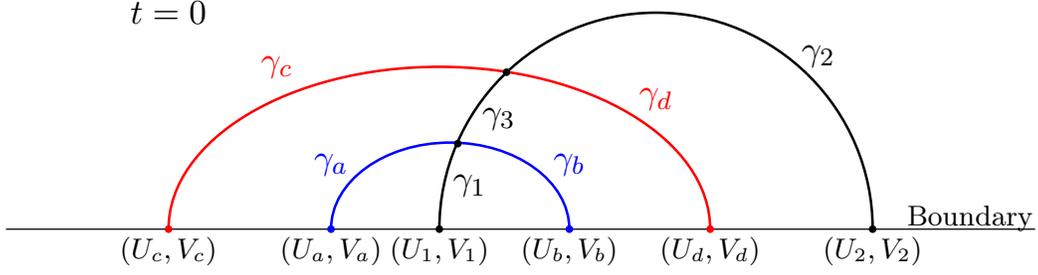
\begin{figure}[t]
    \centering
     
\begin{tikzpicture}[scale=1.2]

\draw (-3.5, 2) node {$t=0$};
\draw (-5,0) -- (4.5,0);
\draw (3.9,.1) node {\footnotesize Boundary};

\draw[thick] (-1,0) arc[start angle=180,delta angle=-180,x radius=2cm, y radius=2cm];
\draw [blue,thick] (-2,0) arc[start angle=180,delta angle=-180,x radius=1.1cm, y radius=.8cm];
\draw [red,thick] (-3.5,0) arc[start angle=180,delta angle=-180,x radius=2.5cm, y radius=1.5cm];

\draw (-3.5,-.2) node {\footnotesize $(U_c,V_c)$};
\draw (-2,-.2) node {\footnotesize $(U_a,V_a)$};
\draw (-1,-.2) node {\footnotesize $(U_1,V_1)$};
\draw (.2,-.2) node {\footnotesize $(U_b,V_b)$};
\draw (1.5,-.2) node {\footnotesize $(U_d,V_d)$};
\draw (3,-.2) node {\footnotesize $(U_2,V_2)$};

\fill [red] (-3.5,0) circle (1pt);
\fill [blue] (-2,0) circle (1pt);
\fill  (-1,0) circle (1pt);
\fill [blue] (.2,0) circle (1pt);
\fill [red] (1.5,0) circle (1pt);
\fill (3,0) circle (1pt);

\fill (-.83,.79) circle (1pt);
% \draw (-1.55,1) node {\footnotesize $(U_{3,ab},V_{3,ab})$};

\fill (-.38,1.45) circle (1pt);
% \draw (-1, 1.7) node {\footnotesize $(U_{3,cd},V_{3,cd})$};

\draw [red] (-2.5,1.5) node {$\gamma_c$};
\draw [red] (1,1.2) node {$\gamma_d$};
\draw [blue] (-2,.6) node {$\gamma_a$};
\draw [blue] (.2,.6) node {$\gamma_b$};
\draw (-.72,.4) node {$\gamma_1$};
\draw (-.45,1.02) node {$\gamma_3$};
\draw (2.5,1.6) node {$\gamma_2$};

\end{tikzpicture}
    \caption{For visual simplicity, we show the configuration projected into a time slice. We start with an HRT surface, shown in black and given by $\gamma=\gamma_1+\gamma_2+\gamma_3$. We then add two constrained geodesics, one in blue ($\gamma_a+\gamma_b$) and one in red ($\gamma_c+\gamma_d$), which each intersect the HRT surface.}
    \label{fig:2congeos}
\end{figure}

We now generalize the network of Figure \ref{fig:new_config} by adding an additional constrained HRT surface as shown in figure  \ref{fig:2congeos}, and which is associated with the links $\gamma_c,\gamma_d$.   We again start with an HRT surface $\gamma$, which is now taken to be anchored to the boundary at $(U_1,V_1)$ and $(U_2,V_2)$. We then add two constrained geodesics: one anchored to $(U_a,V_a)$ and $(U_b,V_b)$, and the other anchored to $(U_c,V_c)$ and $(U_d,V_d)$. See Figure \ref{fig:2congeos}. The two intersection points  divide the original HRT-surface $\gamma$ into the three segments: $\gamma_1$, $\gamma_3$, and $\gamma_2$. 

Since the network of figure \ref{fig:2congeos} contains the network of figure \ref{fig:new_config} as a sub-network, the results of section \ref{sec:halfgeo-results} yield 
\begin{equation}\label{eq:0comm1}
    \{A_{\gamma_a},A_{\gamma_b}\}=\{A_{\gamma_{a,b}},A_{\gamma_1}\}=\{A_{\gamma_{a,b}},A_{\gamma_2\cup\gamma_3}\}=\{A_{\gamma_1},A_{\gamma_2\cup\gamma_3}\}=0,
\end{equation}\label{eq:0comm2}
and, analogously, 
\begin{equation}
    \{A_{\gamma_c},A_{\gamma_d}\}=\{A_{\gamma_{c,d}},A_{\gamma_2}\}=\{A_{\gamma_{c,d}},A_{\gamma_1\cup\gamma_3}\}=\{A_{\gamma_2},A_{\gamma_1\cup\gamma_3}\}=0.
\end{equation}
As a result, only three classes of commutators remain for direct computation:
\begin{equation}
    \{A_{\gamma_{a,b,c,d}},A_{\gamma_3}\}, \{A_{\gamma_{a,b}},A_{\gamma_{c,d}}\}, \text{ and } \{A_{\gamma_{1,2}},A_{\gamma_3}\}.
\end{equation}
Note that we can also obtain $\{A_{\gamma_1},A_{\gamma_2}\}$ from 
$\{A_{\gamma_{1,2}},A_{\gamma_3}\}$ using the final commutator in  \eqref{eq:0comm1}.

Our present goal is not to obtain a full analysis of this network, but rather only to find a case where commutators fail to vanish.  As a result, we will 
focus on the first class of commutators, where we will indeed find a non-vanishing example.

Without loss of generality, we take $U_c<U_a<U_1<U_b<U_d<U_2$, and hence $V_c>V_a>V_1>V_b>V_d>V_2$ to ensure spacelike separation of the anchor points. We will calculate $\{A_{\gamma_a},A_{\gamma_3}\}$. For simplicity, we rewrite the commutator as
\begin{equation}
    \{A_{\gamma_a},A_{\gamma_3}\}=\{A_{\gamma_a},A_{\gamma_1\cup\gamma_3}-A_{\gamma_1}\}=\{A_{\gamma_a},A_{\gamma_1\cup\gamma_3}\},
\end{equation}
where we used $\{A_{\gamma_a},A_{\gamma_1}\}=0$ from \eqref{eq:0comm1}. The functional derivatives of $A_{\gamma_a}$ and $A_{\gamma_1\cup\gamma_3}$ with respect to $\sigma_{U_0}(U)$ are given in Eq.~\eqref{eq:delAgammaa}, respectively. One can also find the analogous expressions for functional derivatives with respect to $\sigma_{V_0}(V)$.

Combining these results with the effective $\sigma$-commutators in  \eqref{eq:effcommU} and \eqref{eq:effcommV}, we find
\begin{equation}\label{eq:a2c}
    \begin{split}
        \{A_{\gamma_a},A_{\gamma_3}\} =& \frac{3}{8}\chi^{(U)}_{21ac} + \frac{1}{8}\chi^{(U)}_{bd21} - \frac{1}{4}C\bigg(\chi^{(U)}_{21ac} + \chi^{(U)}_{bd21}\bigg) \\
        &-\frac{3}{8}\chi^{(V)}_{21ac} - \frac{1}{8}\chi^{(V)}_{bd21} + \frac{1}{4}C\bigg(\chi^{(V)}_{21ac} + \chi^{(V)}_{bd21}\bigg),
    \end{split}
\end{equation}
where we define the cross ratios
\begin{equation}
    \chi^{(U)}_{ijkl} = \frac{(u_i-u_j)(u_k-u_l)}{(u_i-u_k)(u_j-u_l)} \,\,\,\,\text{ and } \,\,\,\,\chi^{(V)}_{ijkl} = \frac{(v_i-v_j)(v_k-v_l)}{(v_i-v_k)(v_j-v_l)}.
\end{equation}
In \eqref{eq:a2c}, $C$ is given in Eq.~\eqref{eq:C}.  Clearly, this result does not vanish in general, though it does vanish in the limit where $(U_c,V_c)$ and $(U_d,V_d)$ approach $(U_2,V_2)$ so that $\gamma_c,\gamma_d$ recede to infinity and our current network reduces to the one studied previously in section \ref{sec:halfgeo-results}. Similar calculations show that other link-area commutators also generally fail to vanish, including commutators between links on different constrained geodesics (e.g. $\gamma_a$ and $\gamma_c$). 

\section{Link-area functional derivatives for the cross section network}\label{sec:func-derivs-app}
In this section we record the functional derivatives for the link areas in the (entanglement wedge) cross section network. First, we find the functional derivatives of the cross ratios $\chi_u$ and $\chi_v$; $\chi_u$ and $\chi_v$ are given in Eq.~\eqref{eq:chiuv2}. Their functional derivatives are hence
\begin{equation}
    \begin{split}\label{eq:delchiu}
        \frac{\delta \chi_u}{\sigma_{U_0}(U)}=& \chi_u e^{-2\sigma_{U_0}(U)}\bigg[ \frac{2}{u_{b2}-u_{a1}}\Theta(U_{b2}-U)\Theta(U-U_{a1}) \\
        &+\frac{2}{u_{a2}-u_{b1}}\Theta(U_{a2}-U)\Theta(U-U_{b1}) - \frac{2}{u_{a2}-u_{a1}}\Theta(U_{a2}-U)\Theta(U-U_{a1}) \\
        &-\frac{2}{u_{b2}-u_{b1}}\Theta(U_{b2}-U)\Theta(U-U_{b1}) \bigg],
    \end{split}
\end{equation}
\begin{equation}
    \begin{split}\label{eq:delchiv}
        \frac{\delta \chi_v}{\sigma_{V_0}(V)}=& \chi_v e^{-2\sigma_{V_0}(V)}\bigg[ \frac{2}{v_{b2}-v_{a1}}\Theta(V_{b2}-V)\Theta(V-V_{a1}) \\
        &+\frac{2}{v_{a2}-v_{b1}}\Theta(V_{a2}-V)\Theta(V-V_{b1}) - \frac{2}{v_{a2}-v_{a1}}\Theta(V_{a2}-V)\Theta(V-V_{a1}) \\
        &-\frac{2}{v_{b2}-v_{b1}}\Theta(V_{b2}-V)\Theta(V-V_{b1}) \bigg],
    \end{split}
\end{equation}
and using these and Eq.~\eqref{eq:delAewcs}, the functional derivative of the cross section area $A_{CS}$ immediately follows
\begin{equation}\label{eq:delACS-full}
\begin{split}
    \frac{\delta A_{CS}}{\sigma_{U_0}(U)}=& -\frac{1}{4\sqrt{1-\chi_u}} e^{-2\sigma_{U_0}(U)}\bigg[ \frac{2}{u_{b2}-u_{a1}}\Theta(U_{b2}-U)\Theta(U-U_{a1}) \\
    &+\frac{2}{u_{a2}-u_{b1}}\Theta(U_{a2}-U)\Theta(U-U_{b1}) - \frac{2}{u_{a2}-u_{a1}}\Theta(U_{a2}-U)\Theta(U-U_{a1}) \\
    &-\frac{2}{u_{b2}-u_{b1}}\Theta(U_{b2}-U)\Theta(U-U_{b1}) \bigg],
\end{split}
\end{equation}
and similarly for the functional derivative with respect to $\hat{\sigma}_{V_0}(V)$. Next, we wish to find the functional derivatives of the half-infinite link areas, as given in Equations \eqref{eq:Aa1}-\eqref{eq:Ab2}, with respect to $\sigma_{U_0}(U)$:
\begin{equation}\label{eq:delAa1}
    \begin{split}
        \frac{\delta A_{a1}}{\delta \sigma_{U_0}(U)} =& \delta(U-U_{a1}) + e^{-2\sigma_{U_0}(U)}\bigg[ \frac{1}{2(u_{a2}-u_{b1})}\Theta(U_{a2}-U)\Theta(U-U_{b1}) \\
        &+\frac{1}{2(u_{b2}-u_{a2})}\Theta(U_{b2}-U)\Theta(U-U_{a2}) \\
        &- \frac{1}{2(u_{a1}-u_{b1})}\Theta(U_{a1}-U)\Theta(U-U_{b1}) \\
        &-\frac{1}{2(u_{b2}-u_{a1})}\Theta(U_{b2}-U)\Theta(U-U_{a1}) \\
        &-\frac{1}{u_{a2}-u_{a1}}\Theta(U_{a2}-U)\Theta(U-U_{a1}) \bigg],
    \end{split}
\end{equation}

\begin{equation}
    \begin{split}
        \frac{\delta A_{a2}}{\delta \sigma_{U_0}(U)}=& \delta(U-U_{a2}) + e^{-2\sigma_{U_0}(U)}\bigg[ \frac{1}{2(u_{a1}-u_{b1})}\Theta(U_{a1}-U)\Theta(U-U_{b1}) \\
        &+\frac{1}{2(u_{b2}-u_{a1})}\Theta(U_{b2}-U)\Theta(U-U_{a1}) \\
        &- \frac{1}{2(u_{a2}-u_{b1})}\Theta(U_{a2}-U)\Theta(U-U_{b1}) \\
        &-\frac{1}{2(u_{b2}-u_{a2})}\Theta(U_{b2}-U)\Theta(U-U_{a2}) \\
        &-\frac{1}{u_{a2}-u_{a1}}\Theta(U_{a2}-U)\Theta(U-U_{a1}) \bigg],
    \end{split}
\end{equation}

\begin{equation}\label{eq:delAb1}
    \begin{split}
        \frac{\delta A_{b1}}{\delta \sigma_{U_0}(U)}=&\delta(U-U_{b1}) + e^{-2\sigma_{U_0}(U)}\bigg[ \frac{1}{2(u_{b2}-u_{a1})}\Theta(U_{b2}-U)\Theta(U-U_{a1}) \\
        &+\frac{1}{2(u_{b2}-u_{a2})}\Theta(U_{b2}-U)\Theta(U-U_{a2}) \\
        &- \frac{1}{2(u_{a1}-u_{b1})}\Theta(U_{a1}-U)\Theta(U-U_{b1}) \\
        &-\frac{1}{2(u_{a2}-u_{b1})}\Theta(U_{a2}-U)\Theta(U-U_{b1}) \\
        &-\frac{1}{u_{b2}-u_{b1}}\Theta(U_{b2}-U)\Theta(U-U_{b1}) \bigg],
    \end{split}
\end{equation}

\begin{equation}\label{eq:delAb2}
    \begin{split}
        \frac{\delta A_{b2}}{\delta \sigma_{U_0}(U)}=&\delta(U-U_{b2}) + e^{-2\sigma_{U_0}(U)}\bigg[ \frac{1}{2(u_{a1}-u_{b1})}\Theta(U_{a1}-U)\Theta(U-U_{b1}) \\
        &+\frac{1}{2(u_{a2}-u_{b1})}\Theta(U_{a2}-U)\Theta(U-U_{b1}) \\
        &- \frac{1}{2(u_{b2}-u_{a1})}\Theta(U_{b2}-U)\Theta(U-U_{a1}) \\
        &-\frac{1}{2(u_{b2}-u_{a2})}\Theta(U_{b2}-U)\Theta(U-U_{a2}) \\
        &-\frac{1}{u_{b2}-u_{b1}}\Theta(U_{b2}-U)\Theta(U-U_{b1}) \bigg],
    \end{split}
\end{equation}
and analogously for functional derivatives with respect to $\hat{\sigma}_{V_0}(V)$. Also useful are the functional derivatives for the half-infinite link areas as expressed in the simpler form given by Eq.~\eqref{eq:Ab1-form} for $A_{b1}$, with analogous expressions for the other link areas. We thus have
\begin{equation}\label{eq:delAb1-form}
    \begin{split}
        \frac{\delta A_{b1}}{\delta \sigma_{U_0}(U)}=&\delta(U-U_{b1}) + \frac{\delta f_{b1}(\chi_u,\chi_v)}{\sigma_{U_0}(U)} \\
        &+ e^{-2\sigma_{U_0}(U)}\bigg[ \frac{1}{u_{a2}-u_{a1}}\Theta(U_{a2}-U)\Theta(U-U_{a1}) \\
        &-\frac{1}{u_{a1}-u_{b1}}\Theta(U_{a1}-U)\Theta(U-U_{b1}) \\
        &- \frac{1}{u_{a2}-u_{b1}}\Theta(U_{a2}-U)\Theta(U-U_{b1}) \bigg]
    \end{split}
\end{equation}
and similarly for the functional derivative with respect to $\hat{\sigma}_{V_0}(V)$. 

\section{Failure of HRT area commutation for a timelike cross section}\label{sec:timelikeCS-app}
In considering the cross section network of Section \ref{sec:ewcs}, we restricted to the case where the cross section $\gamma_{CS}$ is spacelike. We now briefly discuss the case when $\gamma_{CS}$ is timelike. This can be achieved if we take, for instance, the anchor point ordering $U_{b1}<U_{a1}<U_{b2}<U_{a2}$ and $V_{b1}>V_{a1}>V_{b2}>V_{a2}$. We will consider in detail the commutator between $A_{\gamma_{b1}\cup \gamma_{b2}}$, i.e. the area of HRT surface $\gamma_{b1}\cup \gamma_{b2}$, and the cross section area $A_{CS}$. A similar result will hold for the commutator between $A_{\gamma_{a1}\cup \gamma_{a2}}$ and $A_{CS}$. In the case where $\gamma_{CS}$ is spacelike, we expected this commutator to vanish because the flow induced by the HRT area leaves $\gamma_{CS}$ unaffected. We indeed showed this was true via an explicit calculation. However, when $\gamma_{CS}$ is timelike, we do expect it to be affected by the flow induced by the HRT areas in the network. We show this explicitly below.

We begin with a calculation of $\{\chi_u,A_{\gamma_{b1}\cup \gamma_{b2}}\}$. As in all previous calculations, we perform this calculation by integrating over the functional derivatives of the area operators and the effective $\sigma$-commutators in Equations \eqref{eq:effcommU} and \eqref{eq:effcommV}. The $\chi_u$ functional derivative is given by Eq.~\eqref{eq:delchiu}, and the HRT area functional derivative can be found either directly (by differentiating the HRT area) or by adding $\frac{\delta A_{b1}}{\sigma_{U_0}(U)}$ and $\frac{\delta A_{b2}}{\sigma_{U_0}(U)}$ as given in Equations \eqref{eq:delAb1} and \eqref{eq:delAb2}, respectively. We find
\begin{equation}
    \{\chi_u,A_{\gamma_{b1}\cup \gamma_{b2}}\} = \frac{12\pi}{c}(1-\chi_u).
\end{equation}
As explained in Section \ref{sec:ewcs-results}, commutators with $\chi_v$ will be the same up to a sign and replacement of $\chi_u$ with $\chi_v$, yielding
\begin{equation}
    \{\chi_v,A_{\gamma_{b1}\cup \gamma_{b2}}\} = -\frac{12\pi}{c}(1-\chi_v).
\end{equation}
We can now use Eq.~\eqref{eq:delAewcs} to calculate the commutator with $A_{CS}$. We find
\begin{equation}
    \{A_{CS},A_{\gamma_{b1}\cup \gamma_{b2}}\} = -\frac{3\pi}{c} \frac{\sqrt{1-\chi_u}}{\chi_u} + \frac{3\pi}{c}\frac{\sqrt{1-\chi_v}}{\chi_v}.
\end{equation}
A similar result holds for $\{A_{CS},A_{\gamma_{a1}\cup \gamma_{a2}}\}$. As expected, this does not vanish.

\bibliographystyle{jhep}
 	\cleardoublepage

\renewcommand*{\bibname}{References}

\bibliography{CGN}

\end{document}